\documentstyle[12pt,axodraw]{article}
\begin{document}
\newcommand{\bfig}{\begin{center}\begin{picture}}
\newcommand{\efig}[1]{\end{picture}\\{\small #1}\end{center}}
\newcommand{\flin}[2]{\ArrowLine(#1)(#2)}
\newcommand{\wlin}[2]{\DashLine(#1)(#2){2.5}}
\newcommand{\zlin}[2]{\DashLine(#1)(#2){5}}
\newcommand{\glin}[3]{\Photon(#1)(#2){2}{#3}}
\newcommand{\lin}[2]{\Line(#1)(#2)}
\newcommand{\noi}{\noindent}
\newcommand{\sof}{\SetOffset}
\newcommand{\bmip}[2]{\begin{minipage}[t]{#1pt}\bfig(#1,#2)}
\newcommand{\emip}[1]{\efig{#1}\end{minipage}}
\newcommand{\putk}[2]{\Text(#1)[r]{$p_{#2}$}}
\newcommand{\putp}[2]{\Text(#1)[l]{$p_{#2}$}}
\newcommand{\bq}{\begin{equation}}
\newcommand{\eq}{\end{equation}}
\newcommand{\bqa}{\begin{eqnarray}}
\newcommand{\eqa}{\end{eqnarray}}
\newcommand{\nl}{\nonumber \\}
\newcommand{\eqn}[1]{eq. (\ref{#1})}
\newcommand{\eqs}[1]{eqs. (\ref{#1})}
\newcommand{\ibidem}{{\it ibidem\/},}
\newcommand{\vpb}{}
\newcommand{\p}[1]{{\scriptstyle{\,(#1)}}}
\newcommand{\gev}{\mbox{GeV}}
\newcommand{\mev}{\mbox{MeV}}
\setcounter{section}{0}
\setcounter{subsection}{0}
\setcounter{paragraph}{0}
\setcounter{subparagraph}{0}
\setcounter{equation}{0}
\setcounter{figure}{-1}
\setcounter{table}{0}
\setcounter{footnote}{0}
\setcounter{mpfootnote}{0}
\title{
\vspace{-4cm}
\begin{flushright}
{\large ~~~ } \\
\end{flushright}
\vspace{2.7cm}
{\tt NEXTCALIBUR} - A four-fermion generator
for electron-positron collisions
     }
\author{
%\thanks{email address: Roberto.pittau@to.infn.it}
        {\bf F. A. Berends}\\
        Instituut Lorentz, University of Leiden, P. O. Box 9506,\\
        2300 RA Leiden, The Netherlands\\\\
        {\bf C.~G.~Papadopoulos}\\
        Institute of Nuclear Physics, NCSR 'Democritos',\\
        15310 Athens, Greece\\\\
          and\\\\
        {\bf R.~Pittau}\\
        Dipartimento di Fisica Teorica, 
Universit\`a di Torino, Italy\\
INFN, Sezione di Torino, Italy}
\maketitle
\thispagestyle{empty}
\begin{abstract}
\end{abstract}
A fully massive Monte Carlo program
to compute all four-fermion processes in $e^+e^-$ collisions,
including Higgs boson production, is presented. 
Leading higher order effects are discussed and included.

\vspace{1.3cm}

\noindent Pacs: 11.15.-q, 13.10.+q, 13.38.-b, 13.40.-f, 14.70.-e
\clearpage
\section*{PROGRAM SUMMARY}

\noi{\it Title of program:\/} {\tt NEXTCALIBUR}.\\

\noi{\it Program obtainable from:\/} R.~Pittau, 
Dipartimento di Fisica Teorica, Universit\`a di Torino,
Via Giuria 1, I-10125 Torino-Italy.\\

\noi{\it Licensing provisions:\/} none.\\

\noi{\it Computer for which the program is designed and others on which
it has been tested:\/} DIGITAL--ALPHA (double and quadruple precision) 
and HP (double precision only).\\

\noi{\it Operating system under which the program has been tested:\/} UNIX.\\

\noi{\it Programming language used:\/} {\tt FORTRAN 90}.\\

\noi{\it Memory required:\/} about 500kbytes.\\

\noi{\it number of bits per word:\/} 32.\\

\noi{\it Subprograms used:\/} none.\\

\noi{\it Number of lines in distributed program:\/} 10274.\\

\noi{\it Keywords:\/}
decaying vector-boson production, all massive four-fermion processes, 
electroweak background, initial state $p_t$ dependent 
QED radiation, recursive Dyson-Schwinger 
equations, multi-channel Monte Carlo approach.\\

\noi{\it Nature of physical problem\/}

\noi Four-fermion production is and will be investigated at $e^+ e^-$ colliders
in a wide range of energies. Fermion masses have to be included
when studying Higgs boson production and processes with forward scattered
electrons. A Monte Carlo program being able to take into account both
signal and background electroweak diagrams for four-fermion
processes, without neglecting fermion masses, is therefore required. 
Higher order effects, such as the QED radiation evaluated at the proper scale,
have also to be included, together with the possibility of 
generating photons with a non vanishing $p_t$ spectrum. 
For $t$-channel dominated processes, the correct running 
of $\alpha_{EM}$ must be implemented.\\

\noi{\it Method of solution\/}

\noi An event generator is the most suitable choice for a program to be able
to deal with the above physical problems.
For the matrix element evaluation, recursive Dyson-Schwinger 
equations that express the $n$-point Green's functions
in terms of the $1-,2-,\ldots,(n-1)-$point functions are used.
The Monte Carlo integration is performed by using 
an efficient self-optimizing multi-channel approach.\\

\noi{\it Typical running time:\/} about 250 weighted events 
     per second,
     in double precision on DIGITAL--ALPHA/21164 DS20 (500MHz), 
     depending on the chosen physical process.\\

\noi{\it Unusual features of the program:\/} by changing compilation
procedure, the program can run both in double and quadruple precision.
\section*{LONG WRITE-UP}
\section{Introduction}
Electron-positron collisions have provided an ideal testing ground
for the Standard Model. In recent years LEP2 has been instrumental in
this.
Future colliders may take over this role. At these high energies, final
states
will be produced containing unstable particles, like a single $W$-boson,
a Higgs boson or pairs of vector bosons. The net result of this will be
a four-fermion final state. It is then not surprising that event
generators
producing four fermions are in demand. The situation at the start of
LEP2 has been described, amongst others, in the 1996 Yellow Report
~\cite{lep2} and the present status in the 2000 LEP2 Yellow Report
~\cite{wshop}.

  It has been our aim to provide an event generator, which can produce
all possible electroweak four-fermion final states in all possible
configurations. Moreover, its accuracy should match the experimental
requirements as much as is feasible.

  In the early LEP2 days an event generator like {\tt EXCALIBUR} 
~\cite{exca} fulfilled this purpose to a certain extent by producing
efficiently all possible massless four-fermion final states. At present,
the above aim can only be strived for by
\begin{itemize}
\item including fermion masses,
\item improving the treatment of QED radiation,
\item taking into account scale-dependent corrections and higher order
      contributions related to unstable particles.
\end{itemize}

  In the first place, fermion masses are needed both for a description
of 
all kinematical regions (e.g. forward scattering) and for the inclusion
of Higgs particles. 

  In the second place, the accuracy should be improved by extending the
treatment of QED initial state radiation in two ways. A $p_t$ effect of
emitted photons should be generated and a realistic scale for
$t$-channel
dominated processes should be built in.

  Thirdly, the accuracy of the generator can also be increased when
large
fermion-loop effects on vector boson propagators are taken into account.

  The incorporation of these three effects into one event generator led
to a completely new program, {\tt NEXTCALIBUR}. It is the purpose of
this paper to describe the program and the built-in physics treatments.
Section \ref{meva} refers to the new recursive matrix element calculation, 
section \ref{integ} summarizes the phase space generation, whereas 
the following
two sections describe the above mentioned accuracy improving physics
effects. Then, sections \ref{strupro}-\ref{tro} 
are devoted to the actual program, i.e.
its structure, compiling instructions, input and test run output.
Finally, two appendices clarify the multiperipheral phase space 
generation and the adopted fermion-loop scheme.

  For those readers, who are interested in the specific physics results
of {\tt NEXTCALIBUR} and some comparisons to other evaluations we refer
to ref.~\cite{hep}.
\section{The Matrix Element evaluation \label{meva}}
The algorithm for the computation of the matrix element is based on the Dyson-Schwinger equations, a set of recursive
equations that express the $n$-point Green's functions
in terms of the $1-,2-,\ldots,(n-1)-$point functions.
For a detailed presentation of the algorithm as well as its
{\tt FORTRAN } implementation, {\tt HELAC}, we refer to ref.~\cite{helac}.
Here we would like to summarize the main advantages of it:

\begin{enumerate}
\item Scattering amplitudes involving any Standard Model interaction can be treated,
notably the full Higgs sector.
\item Masses of particles are fully considered.
\item Both unitary and Feynman gauges are fully implemented.
\item Quadruple as well as arbitrary precision arithmetic is available,
in order to study special kinematical configurations.
\end{enumerate}

One important point we would like to underline here is that the
algorithm makes use of the chiral representation for fermions and therefore
it is more or less equivalent in efficiency to the method used by {\tt EXCALIBUR}~\cite{exca} for massless fermions.

Because of its modular structure, the algorithm can easily incorporate
higher order contributions to the scattering amplitude, and in fact several
such contributions have been included.
The user has to specify the corresponding flags and input parameters, 
as described is section \ref{input}, in order to study these 
higher order effects. 

\section{The integration strategy \label{integ}}
The integration strategy in {\tt NEXTCALIBUR} is the same adopted 
in ref.~\cite{exca}, namely a multi-channel self-optimizing 
~\cite{wopt} approach.

All possible peaking structures of the integrand
are taken into account by the 19 different kinematical channels 
(plus momenta permutations) in fig. 1. The conventions to read off the
peaking structure are as follows.
Fermionic lines have an arrow, wavy lines represent photons 
while dashed lines are massive gauge bosons. Solid lines connect topological 
equivalent points: a $t$-channel solid line means isotropic angular 
distribution between the connected fermions, while an $s$-channel line 
only gives rise to an $s$ dependent behaviour, without affecting 
the peaking structure. 

The difference with the channels of ref.~\cite{exca} is that now
all fermions are taken to be massive.
\clearpage
\begin{center}
\begin{picture}(390,0)
%--------------------------------------------------
\sof(0,0)
\flin{15,20}{0,0} \flin{0,40}{15,20} \lin{15,20}{30,20}
\flin{45,0}{30,20} \flin{30,20}{45,40}
\glin{37.5,10}{30,-30}{6} \flin{30,-30}{45,-10}
\flin{45,-50}{30,-30}
\Text(0,0)[r]{\small 2} \Text(0,40)[r]{\small 1}
\Text(45,40)[l]{\small 3} \Text(45,0)[l]{\small 4}
\Text(45,-10)[l]{\small 5} \Text(45,-50)[l]{\small 6}
\Text(18,-65)[t]{\tt ANNIHI1}
%--------------------------------------------------
\sof(80,0)
\flin{15,20}{0,0} \flin{0,40}{15,20} \lin{15,20}{30,20}
\flin{45,0}{30,20} \flin{30,20}{45,40}
\wlin{37.5,10}{30,-30} \flin{30,-30}{45,-10}
\Text(30,-10)[r]{{\tt M}}
\flin{45,-50}{30,-30}
\Text(0,0)[r]{\small 2} \Text(0,40)[r]{\small 1}
\Text(45,40)[l]{\small 3} \Text(45,0)[l]{\small 4}
\Text(45,-10)[l]{\small 5} \Text(45,-50)[l]{\small 6}
\Text(18,-64)[t]{\tt ANNIHI2(M)}
%--------------------------------------------------
\sof(160,0)
\flin{0,40}{11,40} \lin{11,40}{33,40} \flin{33,40}{45,40}
\glin{22,40}{22,0}{6}
\flin{11,0}{0,0} \lin{33,0}{11,0} \flin{45,0}{33,0}
\glin{33,0}{30,-30}{6} \flin{30,-30}{45,-10}
\flin{45,-50}{30,-30}
\Text(0,0)[r]{\small 2} \Text(0,40)[r]{\small 1}
\Text(45,40)[l]{\small 3} \Text(45,0)[l]{\small 4}
\Text(45,-10)[l]{\small 5} \Text(45,-50)[l]{\small 6}
\Text(18,-65)[t]{\tt BREMB1}
%--------------------------------------------------
\sof(240,0)
\flin{0,40}{11,40} \lin{11,40}{33,40} \flin{33,40}{45,40}
\glin{22,40}{22,0}{6}
\flin{11,0}{0,0} \lin{33,0}{11,0} \flin{45,0}{33,0}
\wlin{33,0}{30,-30} \flin{30,-30}{45,-10}
\Text(28,-13)[r]{{\tt M}}
\flin{45,-50}{30,-30}
\Text(0,0)[r]{\small 2} \Text(0,40)[r]{\small 1}
\Text(45,40)[l]{\small 3} \Text(45,0)[l]{\small 4}
\Text(45,-10)[l]{\small 5} \Text(45,-50)[l]{\small 6}
\Text(18,-64)[t]{\tt BREMB2(M)}
%--------------------------------------------------
\sof(320,0)
\flin{0,40}{11,40} \lin{11,40}{33,40} \flin{33,40}{45,40}
\glin{22,40}{22,0}{6}
\flin{11,0}{0,0} \lin{33,0}{11,0} \flin{45,0}{33,0}
\glin{11,0}{8,-30}{6} \flin{8,-30}{23,-10}
\flin{23,-50}{8,-30}
\Text(0,0)[r]{\small 2} \Text(0,40)[r]{\small 1}
\Text(45,40)[l]{\small 3} \Text(45,0)[l]{\small 4}
\Text(23,-10)[l]{\small 5} \Text(23,-50)[l]{\small 6}
\Text(18,-65)[t]{\tt BREMF1}
%--------------------------------------------------
\sof(0,-140)
\flin{0,40}{11,40} \lin{11,40}{33,40} \flin{33,40}{45,40}
\glin{22,40}{22,0}{6}
\flin{11,0}{0,0} \lin{33,0}{11,0} \flin{45,0}{33,0}
\wlin{11,0}{8,-30} \flin{8,-30}{23,-10}
\Text(6,-13)[r]{{\tt M}}
\flin{23,-50}{8,-30}
\Text(0,0)[r]{\small 2} \Text(0,40)[r]{\small 1}
\Text(45,40)[l]{\small 3} \Text(45,0)[l]{\small 4}
\Text(23,-10)[l]{\small 5} \Text(23,-50)[l]{\small 6}
\Text(18,-64)[t]{\tt BREMF2(M)}
%--------------------------------------------------
\sof(80,-140)
\flin{0,40}{11,40} \lin{11,40}{33,40} \flin{33,40}{45,40}
\lin{22,40}{22,0}
\flin{11,0}{0,0} \lin{33,0}{11,0} \flin{45,0}{33,0}
\glin{11,0}{8,-30}{6} \flin{8,-30}{23,-10}
\flin{23,-50}{8,-30}
\Text(0,0)[r]{\small 2} \Text(0,40)[r]{\small 1}
\Text(45,40)[l]{\small 3} \Text(45,0)[l]{\small 4}
\Text(23,-10)[l]{\small 5} \Text(23,-50)[l]{\small 6}
\Text(18,-65)[t]{\tt BREMF3}
%--------------------------------------------------
\sof(160,-140)
\flin{0,40}{11,40} \lin{11,40}{33,40} \flin{33,40}{45,40}
\lin{22,40}{22,0}
\flin{11,0}{0,0} \lin{33,0}{11,0} \flin{45,0}{33,0}
\wlin{11,0}{8,-30} \flin{8,-30}{23,-10}
\Text(6,-13)[r]{{\tt M}}
\flin{23,-50}{8,-30}
\Text(0,0)[r]{\small 2} \Text(0,40)[r]{\small 1}
\Text(45,40)[l]{\small 3} \Text(45,0)[l]{\small 4}
\Text(23,-10)[l]{\small 5} \Text(23,-50)[l]{\small 6}
\Text(18,-64)[t]{\tt BREMF4(M)}
%--------------------------------------------------
\sof(240,-140)
\flin{0,40}{15,20} \flin{15,20}{15,-30} \flin{15,-30}{0,-50}
\glin{15,20}{30,20}{3}
\glin{15,-30}{30,-30}{3}
\flin{45,0}{30,20} \flin{30,20}{45,40}
\flin{30,-30}{45,-10} \flin{45,-50}{30,-30}
\Text(0,-50)[r]{\small 2} \Text(0,40)[r]{\small 1}
\Text(45,40)[l]{\small 3} \Text(45,0)[l]{\small 4}
\Text(45,-10)[l]{\small 5} \Text(45,-50)[l]{\small 6}
\Text(18,-65)[t]{\tt CONVER1}
%--------------------------------------------------
\sof(320,-140)
\flin{0,40}{15,20} \flin{15,20}{15,-30} \flin{15,-30}{0,-50}
\glin{15,20}{30,20}{3}
\wlin{15,-30}{30,-30}
\Text(23,-26)[b]{{\tt M}}
\flin{45,0}{30,20} \flin{30,20}{45,40}
\flin{30,-30}{45,-10} \flin{45,-50}{30,-30}
\Text(0,-50)[r]{\small 2} \Text(0,40)[r]{\small 1}
\Text(45,40)[l]{\small 3} \Text(45,0)[l]{\small 4}
\Text(45,-10)[l]{\small 5} \Text(45,-50)[l]{\small 6}
\Text(18,-64)[t]{\tt CONVER2(M)}
%--------------------------------------------------
\sof(0,-280)
\flin{0,40}{15,20} \flin{15,20}{15,-30} \flin{15,-30}{0,-50}
\wlin{15,20}{30,20}
\Text(23,24)[b]{{\tt M1}}
\wlin{15,-30}{30,-30}
\Text(23,-26)[b]{{\tt M2}}
\flin{45,0}{30,20} \flin{30,20}{45,40}
\flin{30,-30}{45,-10} \flin{45,-50}{30,-30}
\Text(0,-50)[r]{\small 2} \Text(0,40)[r]{\small 1}
\Text(45,40)[l]{\small 3} \Text(45,0)[l]{\small 4}
\Text(45,-10)[l]{\small 5} \Text(45,-50)[l]{\small 6}
\Text(18,-64)[t]{\tt CONVER3(M1,M2)}
%--------------------------------------------------
\sof(80,-280)
\flin{0,40}{11,40} \lin{11,40}{33,40} \flin{33,40}{45,40}
\glin{22,40}{22,10}{4}
\flin{22,10}{45,10} \flin{22,-20}{22,10} \flin{45,-20}{22,-20}
\glin{22,-50}{22,-20}{4}
\flin{11,-50}{0,-50} \lin{33,-50}{11,-50} \flin{45,-50}{33,-50}
\Text(0,-50)[r]{\small 2} \Text(0,40)[r]{\small 1}
\Text(45,40)[l]{\small 3} \Text(45,10)[l]{\small 4}
\Text(45,-20)[l]{\small 5} \Text(45,-50)[l]{\small 6}
\Text(18,-65)[t]{\tt MULTI1}
%--------------------------------------------------
\sof(160,-280)
\flin{0,40}{11,40} \lin{11,40}{33,40} \flin{33,40}{45,40}
\glin{22,40}{22,10}{4}
\flin{22,10}{45,10} \flin{22,-20}{22,10} \flin{45,-20}{22,-20}
\lin{22,-50}{22,-20}
\flin{11,-50}{0,-50} \lin{33,-50}{11,-50} \flin{45,-50}{33,-50}
\Text(0,-50)[r]{\small 2} \Text(0,40)[r]{\small 1}
\Text(45,40)[l]{\small 3} \Text(45,10)[l]{\small 4}
\Text(45,-20)[l]{\small 5} \Text(45,-50)[l]{\small 6}
\Text(18,-65)[t]{\tt MULTI2}
%--------------------------------------------------
\sof(240,-280)
\flin{0,40}{11,40} \lin{11,40}{33,40} \flin{33,40}{45,40}
\lin{22,40}{22,10}
\flin{22,10}{45,10} \flin{22,-20}{22,10} \flin{45,-20}{22,-20}
\lin{22,-50}{22,-20}
\flin{11,-50}{0,-50} \lin{33,-50}{11,-50} \flin{45,-50}{33,-50}
\Text(0,-50)[r]{\small 2} \Text(0,40)[r]{\small 1}
\Text(45,40)[l]{\small 3} \Text(45,10)[l]{\small 4}
\Text(45,-20)[l]{\small 5} \Text(45,-50)[l]{\small 6}
\Text(18,-65)[t]{\tt MULTI3}
%--------------------------------------------------
\sof(320,-280)
\flin{0,15}{15,-5} \flin{15,-5}{0,-25}
\lin{15,-5}{30,-5}
\wlin{30,20}{30,-5}
\wlin{30,-5}{30,-30}
\Text(21,3)[b]{{\tt M1}}
\Text(21,-12)[t]{{\tt M2}}
\flin{45,0}{30,20} \flin{30,20}{45,40}
\flin{30,-30}{45,-10} \flin{45,-50}{30,-30}
\Text(0,-25)[r]{\small 2} \Text(0,15)[r]{\small 1}
\Text(45,40)[l]{\small 3} \Text(45,0)[l]{\small 4}
\Text(45,-10)[l]{\small 5} \Text(45,-50)[l]{\small 6}
\Text(18,-64)[t]{\tt NONAB1(M1,M2)}
%--------------------------------------------------
\sof(0,-420)
\flin{0,40}{11,40} \lin{11,40}{33,40} \flin{33,40}{45,40}
\glin{13,40}{13,-5}{4}
\lin{13,-5}{13,-50}
\wlin{13,-5}{33,-5}
\Text(23,-1)[b]{{\tt M}}
\flin{33,-5}{45,15} \flin{45,-25}{33,-5}
\flin{11,-50}{0,-50} \lin{33,-50}{11,-50} \flin{45,-50}{33,-50}
\Text(0,-50)[r]{\small 2} \Text(0,40)[r]{\small 1}
\Text(45,40)[l]{\small 3} \Text(45,15)[l]{\small 4}
\Text(45,-25)[l]{\small 5} \Text(45,-50)[l]{\small 6}
\Text(18,-64)[t]{\tt NONAB2(M)}
%--------------------------------------------------
\sof(80,-420)
\flin{0,40}{11,40} \lin{11,40}{33,40} \flin{33,40}{45,40}
\lin{13,40}{13,-5}
\lin{13,-5}{13,-50}
\glin{13,-5}{33,-5}{4}
\flin{33,-5}{45,15} \flin{45,-25}{33,-5}
\flin{11,-50}{0,-50} \lin{33,-50}{11,-50} \flin{45,-50}{33,-50}
\Text(0,-50)[r]{\small 2} \Text(0,40)[r]{\small 1}
\Text(45,40)[l]{\small 3} \Text(45,15)[l]{\small 4}
\Text(45,-25)[l]{\small 5} \Text(45,-50)[l]{\small 6}
\Text(18,-65)[t]{\tt NONAB3}
%--------------------------------------------------
\sof(160,-420)
\flin{0,40}{11,40} \lin{11,40}{33,40} \flin{33,40}{45,40}
\lin{13,40}{13,-5}
\lin{13,-5}{13,-50}
\wlin{13,-5}{33,-5}
\Text(23,-1)[b]{{\tt M}}
\flin{33,-5}{45,15} \flin{45,-25}{33,-5}
\flin{11,-50}{0,-50} \lin{33,-50}{11,-50} \flin{45,-50}{33,-50}
\Text(0,-50)[r]{\small 2} \Text(0,40)[r]{\small 1}
\Text(45,40)[l]{\small 3} \Text(45,15)[l]{\small 4}
\Text(45,-25)[l]{\small 5} \Text(45,-50)[l]{\small 6}
\Text(18,-64)[t]{\tt NONAB4(M)}
%--------------------------------------------------
\sof(240,-420)
\flin{0,15}{15,-5} \flin{15,-5}{0,-25}
\lin{15,-5}{30,-5}
\lin{30,20}{30,-5}
\lin{30,-5}{30,-30}
\flin{45,0}{30,20} \flin{30,20}{45,40}
\flin{30,-30}{45,-10} \flin{45,-50}{30,-30}
\Text(0,-25)[r]{\small 2} \Text(0,15)[r]{\small 1}
\Text(45,40)[l]{\small 3} \Text(45,0)[l]{\small 4}
\Text(45,-10)[l]{\small 5} \Text(45,-50)[l]{\small 6}
\Text(18,-65)[t]{\tt RAMB04}
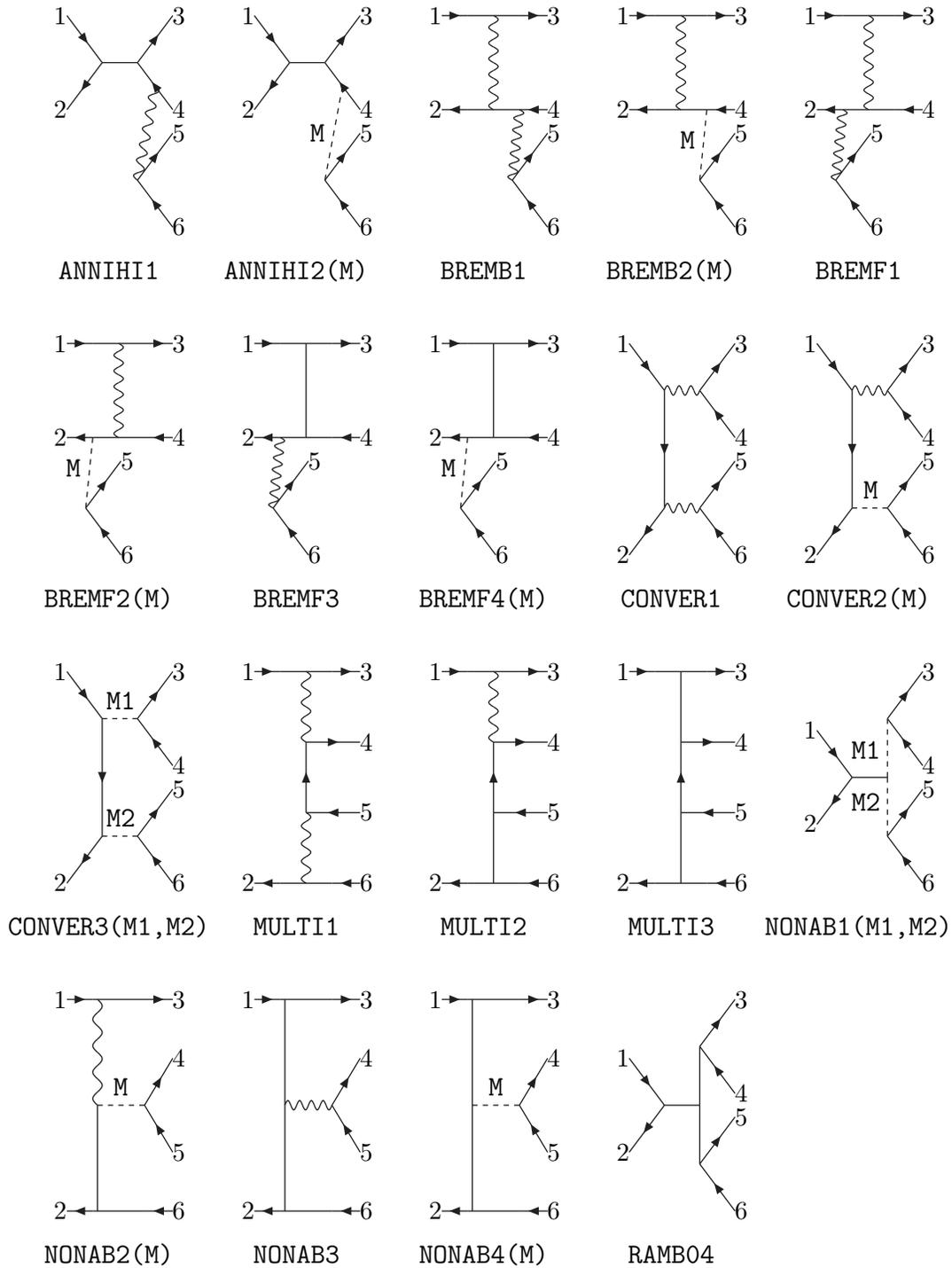
\Text(-80,-100)[]{Figure 1: kinematical diagrams in {\tt NEXTCALIBUR}.}
\label{fig1}
\end{picture}
\end{center}
\clearpage
Furthermore, the leading kinematical structures for Higgs 
boson production 
have been added. They are represented by the channels {\tt NONAB1(M1,M2)} 
and {\tt NONAB4(M)}, where the masses can now also be the Higgs mass. 

Generally speaking, the modification from the old massless to the
new massive channels required only trivial changes. An exception is
represented by the channel {\tt MULTI1}, for which 
we list, in appendix \ref{seca}, the complete generation algorithm.

A last remark is in order. When dealing with final state electrons,
strong numerical cancellations occur, that may
degrade the accuracy of the phase space generation.
In order to overcome this problem a number of tricks are applied 
all over the places in {\tt NEXTCALIBUR}, all derived from
the basic identity 
\bqa
q_0-|\vec{q}| = \frac{m^2_e}{(q_0+|\vec{q}|)}\,,~~~{\rm where}
~~~q^2= m^2_e\,,
~~~q = (q_0,\vec{q})\,.
\eqa

\section{Treatment of the QED radiation \label{sec3}}
In {\tt NEXTCALIBUR} QED radiation is implemented via
the Structure Function formalism, namely by convoluting the 
Born cross section together with QED Initial State Radiators~\cite{isr}.
Two ISR photons are explicitly generated, by using  
$p_t$ dependent Structure Functions~\cite{ptsf}, derived, 
at the first leading logarithmic order, for small values of $p_t$
~\cite{ptsf1}.
Our starting point is the convolution
\bqa
\sigma(s)= \int dx_1\,dx_2\,dc_1\,dc_2\,
\frac{1}{2 \pi}\int_0^{2 \pi} d\phi_1\,
\frac{1}{2 \pi}\int_0^{2 \pi} d\phi_2\,
\Phi(x_1,c_1) \Phi(x_2,c_2)\,\sigma_0(\hat s)\,,
\label{eq0}
\eqa
where $\sigma_0$ is the Born four-fermion cross section,
$\hat s$ the reduced center of mass energy of the event, after
photon emission, $c_{1,2}$ and $\phi_{1,2}$ the cosines of
the directions (in the laboratory frame) and the azimuthal angles 
of the two emitted photons with respect to the incoming particles.
Finally, the Structure Functions are given by 
\bqa
&&\!\!\!\!\!\!\!\!\!\!\!\Phi(x,c,q^2) = \nl
&&\!\frac{{\rm exp}\left\{\frac{1}{2}\beta
\left(\frac{3}{4}-\gamma_E \right)\right\}}{\Gamma
\left(1+\frac{1}{2}\beta\right)} \frac{\alpha}{\pi}  
(1-x)^{\frac{\beta}{2}-1}\left(
\frac{1}{1-c+2\frac{m^2_e}{q^2}} -2 \frac{m^2_e}{q^2}
\frac{1}{(1-c+2\frac{m^2_e}{q^2})^2}
                        \right) \nonumber \nl
&-&\!\frac{\alpha}{2 \pi}(1+x)\left(
   \frac{1}{1-c+2\frac{m^2_e}{q^2}}
  +\frac{1-x}{1+x} \cdot \frac{1}{2}
   -4 \frac{m^2_e}{q^2}
\frac{1}{(1+x)(1-c+2\frac{m^2_e}{q^2})^2}
                            \right) \nonumber \nl
&+&\!\frac{\beta}{8} \frac{\alpha}{2 \pi}
\left[-4(1+x)\ln(1-x)+3(1+x) \ln x -4\frac{\ln (x)}{1-x}-5-x \right] 
\nonumber \nl
&\times&\!                                       \left(
   \frac{1}{1-c+2\frac{m^2_e}{q^2}}
  +\frac{1-x}{1+x} \cdot \frac{1}{2}
   -4 \frac{m^2_e}{q^2} 
\frac{1}{(1+x)(1-c+2\frac{m^2_e}{q^2})^2}
                                       \right)\,,
\label{eq1}
\eqa
with
\bqa
\beta= 2 \frac{\alpha}{\pi} \left(\ln(\frac{q^2}{m^2_e})-1\right)\,.
\label{eq1a}
\eqa
The above formula can be obtained by {\em unfolding}
the strictly collinear Structure Functions, with the replacement
\bqa
\ln(\frac{q^2}{m^2_e})-1~~\to~~ 
\frac{1}{1-c+2\frac{m^2_e}{q^2}} -2 \frac{m^2_e}{q^2}
\frac{1}{(1-c+2\frac{m^2_e}{q^2})^2}
\label{eq2}
\eqa
 for the soft part, and
\bqa
\ln(\frac{q^2}{m^2_e})-1~~\to~~ 
\frac{1}{1-c+2\frac{m^2_e}{q^2}}
+\frac{1-x}{1+x} \cdot \frac{1}{2}
-4 \frac{m^2_e}{q^2}
\frac{1}{(1+x)(1-c+2\frac{m^2_e}{q^2})^2}
\label{eq3}
\eqa
for the hard contributions
\footnote{Note that the replacement of eq. (\ref{eq3}) is 
also performed in the higher order terms of eq. (\ref{eq1}).}.
With this choice, the whole expression gets proportional to
$\ln(\frac{q^2}{m^2_e})-1$, after integrating over $c$, as it should be.
The inclusive QED result is therefore recovered and, 
at the same time, the pattern of the photon radiation is exact 
for small values of $p_t$.

The generation algorithm is as follows.
The initial state electrons, before radiating, have 
momenta given by
\bqa
&&p_1 = (E, \beta E,0,0)\,,~p_2= (E, -\beta E,0,0)\,,
~E= \frac{\sqrt{s}}{2}\,,
~\beta= \sqrt{1-\frac{m^2_e}{E^2}}\,.~~
\label{eq3a}
\eqa
Once $c_{1,2}$ are generated, together with the energy
fractions $x_{1,2}$ and the azimuthal angles $\phi_{1,2}$,
the momenta of the two ISR photons are completely determined
\bqa
&&q_1= E(1-x_1)(1, c_1,s_1  \cos\phi_1,s_1 \sin\phi_1) \,,\nl
&&q_2= E(1-x_2)(1,-c_2,s_2  \cos\phi_2,s_2 \sin\phi_2) \,,~~~~
s_i= \sqrt{1-c^2_i} \,,
\eqa
and the invariant mass, after ISR, is given by
\bqa
\hat s &=& \ell^2 \equiv (\ell_1+\ell_2)^2 \nonumber \nl
       &=& s \left\{ 
             x_1 x_2+\frac{(1-x_1)(1-x_2)}{2}
             \left[c_1 c_2-1-s_1 s_2 \cos(\phi_1-\phi_2)\right]
             \right \}\,, \nl
  \ell_1 &\equiv& p_1-q_1\,,~~~~~\ell_2~\equiv~ p_2-q_2\,.
\label{eq4}
\eqa
We then generate four-fermion events in the center-of-mass frame 
of $\ell$, where the new $x$ direction {\em after} ISR is determined by
the spatial part of the four-vector obtained by boosting 
$\ell_1$ in the center-of-mass frame.
 All generated four-fermion momenta are subsequently boosted 
back to the Laboratory frame.

 In {\tt NEXTCALIBUR} there is also the possibility of generating strictly
collinear radiation. In this case the relevant formulae are
\bqa
\sigma(s) &=& \int dx_1~dx_2~
\Phi(x_1) \Phi(x_2)~\sigma_0(\hat s)\,,\nonumber \nl
\Phi(x,q^2)   &=& 
\frac{{\rm exp}\left\{\frac{1}{2}\beta
\left(\frac{3}{4}-\gamma_E \right)
               \right\}}{\Gamma\left(1+\frac{1}{2}\beta\right)}
 \frac{\beta}{2}(1-x)^{\frac{\beta}{2}-1}
-\frac{\beta}{4}(1+x) \label{eq4a} \\
          &+&\frac{1}{32}\beta^2
\left[-4(1+x)\ln(1-x)+3(1+x) \ln x -4\frac{\ln (x)}{1-x}-5-x \right]\,,
\nonumber
\eqa
with $\beta$ defined in eq. (\ref{eq1a}).
In order to allow particular QED studies, the generation of 
photons in {\tt NEXTCALIBUR} has been kept independent
for each of the two incoming legs. This implies the possibility of 
two different scales $q^2_-$ and $q^2_+$ for the electron 
and positron leg respectively. Mixed situations are also
possible, in which one leg produces a collinear photon, while the other
generates a photon with $p_t$. This can be controlled 
at the level of the input file, as explained in the next sections.
 
A second issue, relevant when studying QED radiation, is the choice of
the $q^2$ to be inserted in the Structure Functions of eq. (\ref{eq1}) 
and/or eq. (\ref{eq4a}). The choice $q^2 \sim s$ is proven 
to reproduce accurately the inclusive four-fermion cross sections, 
at least for $s$-channel dominated processes. 
  For $t$-channel dominated processes a different choice is
more adequate, as studies of certain processes
have shown. When an exact first order QED radiative correction
calculation exists for a $t$-channel dominated process, then the
result can be compared to a Structure Function calculation
with a $q^2$ scale related to the virtuality of the exchanged
$t$-channel photon. With such a $q^2$ value the two kinds of
calculations agree for small angle Bhabha scattering~\cite{sbhab}
and multi-peripheral two-photon processes~\cite{isr2}, where the exact
calculation already exists for some time~\cite{gge}. When no exact
first order QED correction calculation is available, the first order
soft correction may also serve as a guideline to determine $q^2$
~\cite{isr2,isr3}.
In {\tt NEXTCALIBUR}, the choice of the scale is performed automatically,
event by event, according to the selected final state, as shown in Table 1.

\vspace{0.4cm}

\begin{center}
\begin{tabular}{|l||c|c|} \hline 
Final State          & $q^2_-$  & $q^2_+$ \\ \hline \hline
No $e^\pm$ & $s$     & $s$     \\ \hline              
1  $e^-$   & $|t_-|$ & $s$     \\ \hline               
1  $e^+$   & $s$     & $|t_+|$ \\ \hline               
1  $e^-$ and 1  $e^+$  & $|t_-|$    & $|t_+|$ \\ \hline               
2  $e^-$ and 2  $e^+$  & min($|t_-|$)    & min($|t_+|$) \\ \hline
\end{tabular}
\end{center}
{\em Table 1: The choice of the QED scale in {\tt NEXTCALIBUR}.
$q^2_\pm$ are the scales of the incoming $e^\pm$ while
$t_\pm$ represent the $t$-channel invariants obtained
by combining initial and final state $e^\pm$ momenta.
When two combinations are possible, as in the last entry of the table, 
that one with the minimum value of $|t|$ is chosen, event by event.}

\vspace{0.4cm}
When using Structure Functions with a $t$-channel scale,
one faces inefficiencies in the event generation.
To illustrate the problem, we start from a
simple model, namely a single collinear photon generated 
with a distribution given by the leading behaviour of
eq. (\ref{eq4a}), namely
\bqa
\Phi(x) \sim \frac{1}{(1-x)^{(1-z)}}\,,~~~~
z= \frac{\alpha}{\pi} \left(\ln(\frac{|t|}{m^2_e})-1\right)\,.
\label{eq5}
\eqa
The relevant integral is
\bqa
I &=& \int_0^1 dx \int d(PS)_4\, 
\Phi(x)\, |M(t)|^2\,,
\eqa
where $d(PS)_4$ is the four-body phase-space integration element
and $|M(t)|^2$ the
$t$ dependent four-fermion Born matrix element squared.
The usual way to cure the peaking behaviour of $\Phi(x)$
would be performing a change of variable
\bqa
(1-x)= \rho^\frac{1}{z_g}\,,~0<\rho<1~{\rm uniformly}
\label{eq6}
\eqa
so that
\bqa
I &=& \int_0^1 d\rho \int d(PS)_4 \left(-\frac{1}{z_g} \right)\,
(1-x)^{(z-z_g)} \, |M(t)|^2\,,
\label{eq6a}
\eqa
and choosing $z_g = z$, namely
\bqa
z_g= \frac{\alpha}{\pi} \left(\ln(\frac{t_g}{m^2_e})-1\right)
\label{eq7}
\eqa 
with $t_g = |t|$.

\noindent The problem here is that, with our QED model, $x$ has 
to be generated, through eqs. (\ref{eq6})
and (\ref{eq7}), {\em before} knowing $t$. 
The latter is in fact related to the 
subsequent generation of the four-fermion event. 
An initial $t_g$ must therefore be employed, that is
not directly related to the {\em true} generated $t$. 
Therefore, event by event, the numerical 
value of the term 
\bqa
(1-x)^{(z-z_g)} = (1-x)^{\frac{\alpha}{\pi} \ln \frac{|t|}{t_g}}
\label{eq7a}
\eqa
in eq. (\ref{eq6a}) may vary considerably,
leading to Monte Carlo inefficiencies.

\noindent Our strategy is as follows. Since eq. (\ref{eq6a}) 
does not depend on $t_g$, we are free to integrate 
over an extra uniform variable $0 < \rho_0< 1$, that we also use 
to generate $t_g$ distributed as
$1/{t_g}$ between a minimum ($t_g^-$) and a maximum value ($t_g^+= s$)
\bqa
t_g = t_g^- \left(\frac{t_g^+}{t_g^-}\right)^{\rho_0}\,.
\label{eq8}
\eqa
Our integral then becomes
\bqa
I &=& \int_0^1 d\rho_0 
\int_0^1 d\rho \int d(PS)_4 \left(-\frac{1}{z_g} \right)\,
(1-x)^{(z-z_g)} \, |M(t)|^2\,,
\eqa
with $z_g$ and $t_g$ given in eqs. (\ref{eq7}) and (\ref{eq8}).   
We have chosen $t_g$ distributed as 
$1/{t_g}$ because this is the expected behaviour of the {\em true}
$t$ coming from $|M(t)|^2$, when small $t$-channel scales dominate.
However, this alone does not yet prevent a blowing 
of the term in eq. (\ref{eq7}) for a few events.
We therefore split the integration range of $\rho_0$ into $n$
equidistant bins, introduce a set of weights 
$\alpha_i$, normalized such that $\sum_{i=1}^{n}\alpha_i = 1$,
and rewrite
\bqa
I &=& n\,\sum_{i=1}^{n}\,I_i\,, \nl
I_i &=& \int_{\rho_{0i}^-}^{\rho_{0i}^+} d\rho_0 
\int_0^1 d\rho \int d(PS)_4\,\alpha_i\,\left(-\frac{1}{z_g} \right)\,
(1-x)^{(z-z_g)} \, |M(t)|^2\,,\nl
\rho_{0i}^- &=& \frac{i-1}{n}\,,~~~~\rho_{0i}^+ = \frac{i}{n}\,.
\label{eq9}
\eqa
Since $\rho_{0i}^- < \rho_0 < \rho_{0i}^+$ corresponds to
the $i^{th}$ generation interval
\bqa
t_{gi}^-  < t_g < t_{gi}^+ \,,
~~~~t_{gi}^\pm = t_g^-\left(\frac{t_g^+}{t_g^-}\right)^{\rho_{0i}^\pm}\,,
\label{eq10}
\eqa
eq. (\ref{eq9}) allows to weight differently events 
produced in different intervals.
Notice that the set of weights $\{\alpha_i\}$ can depend on $t$. 
A suitable choice is
\bqa
\alpha_i &\equiv& \alpha_i(|t_j|^-) = 
\frac{(1-x)^{\frac{\alpha}{\pi} \ln \frac{t_{gi}^-}{|t_j|^-}}}{
\sum_i (1-x)^{\frac{\alpha}{\pi} \ln \frac{t_{gi}^-}{|t_j|^-}}}\,.
\eqa
The meaning of $|t_j|^-$ in the above formula is as follows.
Once $t_g$ is produced as described, the four-fermion event 
can be generated with a given value of $t$ (the {\em true} t).
The variable $|t|$ can then be binned as done 
for $t_g$ in eq. (\ref{eq10}) and, if $t_{gj}^-  < |t| < t_{gj}^+$,
$|t_j|^-= t_{gj}$.
Then, even by event, the set $\{\alpha_i\}$ changes, compensating
the term in eq. (\ref{eq7a}) and improving the generation efficiency.

Summarizing, the Monte Carlo implementation of the technique is
\begin{itemize}
\item[1)] divide the interval [0,1] in $n$ equidistant bins;
\item[2)] generate $\rho_0$ uniformly in [0,1]\\ 
          $\Rightarrow$ $i$ and $t_{gi}^-$ are known;
\item[3)] generate ISR and the four-fermion event\\ 
          $\Rightarrow$ the event weight $w$, the variable $t$, $j$
          and $|t_j|^-$ are known;
\item[4)] $w \to w\, \cdot n\, \cdot \alpha_i(|t_j|^-)$\,.
\end{itemize}
The algorithm works unchanged when both initial 
state particles radiate, and when photons are generated 
with a non vanishing $p_t$. 
In this latter case, eq. (\ref{eq5}) is replaced
by
\bqa
\Phi(x,c) \sim \frac{1}{(1-x)^{(1-z)}}\,
\frac{1}{1-c+2\frac{m^2_e}{|t|}}\,.
\eqa
The generation of $x$ is as described previously. In addition,
$c$ is generated with distribution 
$$\frac{1}{1-c+2\frac{m^2_e}{t_g}},$$
and the weights $\alpha_i$ modified accordingly, 
in order to compensate the term
\bqa
\frac{1-c+2\frac{m^2_e}{t_g}}{1-c+2\frac{m^2_e}{|t|}}
\eqa
appearing in the event weight.

 A last comment is in order on our QED modelling.
By using $q^2= |t|$ in the Structure Functions, a few events 
are produced with very low values of $\ln(|t|/m^2_e)$, when $|t| \sim m^2_e$.
This means that logarithms are no longer leading, 
breaking down the Structure Function approach. 
Constant terms should in principle be included. 
 On the other hand, low values of $\ln(|t|/m^2_e)$ imply
$\Phi(x) \sim \delta(1-x)$, resulting in a suppression of the QED radiation,
as confirmed by direct calculations~\cite{sbhab,gge}.
 Notice that even implementing the exact form of the
Structure Functions, also valid for $|t| < m^2_e$~\cite{bfact}, 
only includes the 
factorizable part of the missing constants and not all of them. 
In the present version of {\tt NEXTCALIBUR} we follow a different strategy
and introduce a minimum $|t|_{min}$, such that, if
$|t| < |t|_{min}$, the ISR logarithm is always evaluated
at $q^2 = |t|_{min}$
\footnote{We consequently put $t_g^-= |t|_{min}$ in eq. (\ref{eq8}).}.
With $|t|_{min} = 100\,m^2_e$, the minimum value of the logarithm 
still gives physical results, but is small enough to reproduce 
the discussed radiation suppression. We checked the
insensitivity of our results to the choice of $|t|_{min}$
in the range $50\,m^2_e <|t|_{min} < 200\,m^2_e$.
\section{The running of $\alpha_{EM}$ \label{alphar}}
When studying high energy processes, part of the higher order 
corrections can be reabsorbed in the Born approximation
by using the so-called $G_F$ scheme. 
In such a scheme $G_F$, $M_Z$ and $M_W$ are 
input parameters, while the weak mixing angle and 
$\alpha_{EM}$ are derived quantities:
\bqa
s^2_W &=& 1 - {M^2_W}/{M^2_Z}\,, \nl
\alpha_{EM} &=&  \sqrt{2}\,\,\frac{G_F\,M_W^2\,s^2_W}{\pi} \,.
\eqa
In the presence of low $t$-channel scales such an approach fails,
since the choice $\alpha(t \sim 0) \sim 1/137$ is certainly
more appropriate for certain sets of diagrams and/or
kinematical regions.
 The question is therefore how to include consistently the running
of $\alpha_{EM}$ in four-fermion processes with one or more electrons
in the final state.

An exact and field-theoretically consistent solution to this
problem is represented by the 
Fermion-Loop approach of refs.~\cite{passa1}-\cite{passa3}, 
where the whole set of fermion one-loop corrections is
taken into account by computing running couplings 
$g(s)$ and $e(s)$ and re-summed bosonic propagators.

 In the presence of the $W W \gamma$ vertex, also loop mediated 
vertices are required to preserve gauge invariance.
On the contrary, when the $W W \gamma$ coupling is absent, 
the neutral gauge boson vertices, induced by the fermion loop contributions, 
are separately gauge invariant~\cite{passa2}.

 Since the exact expression of the loop-induced vertex functions 
is rather cumbersome, we follow a simplified approach
(Modified Fermion-Loop approach) by
neglecting the separately gauge invariant neutral 
boson vertices and including only the part of the $W W \gamma $ 
loop function necessary to preserve the $U(1)$ gauge invariance.
Besides running couplings,
we use bosonic propagators
%--
\bqa
&&P_w^{\mu\nu}(s)= \left(s-M^2_W(s)\right)^{-1}
 \left(g_{\mu\nu}-\frac{p_\mu p_\nu}{M^2_W(s)}\right)
 \nonumber \\
&&P_z^{\mu\nu}(s)= \left(s-M^2_Z(s)\right)^{-1} 
 \left(g_{\mu\nu}-\frac{p_\mu p_\nu}{M^2_Z(s)}\right)\,,
\label{bosprop}
\eqa
%--
with running boson masses defined as
%--
\bqa
&& M^2_W(s)= \mu_w\frac{g^2(s)}{g^2(\mu_w)}
-g^2(s) [T_W(s)-T_W(\mu_w)]\,\nonumber \\
&& M^2_Z(s)= \mu_z \frac{g^2(s)}{c^2_\theta(s)} 
\frac{c^2_\theta(\mu_z)}{g^2(\mu_z)}
-\frac{g^2(s)}{c^2_\theta(s)}[T_Z(s)-T_Z(\mu_z)]\,. 
\eqa
%--
$T_{W,Z}(s)$ are contributions due to the top quark,
$\mu_{w,z}$ the complex poles of the propagators and
$$ s^2_\theta(s)= \frac{e^2(s)}{g^2(s)}\,,~c^2_\theta(s)= 1-s^2_\theta(s)\,.$$
The explicit form of the running functions $e^2(s),~g^2(s)$ and
$T_{W,Z}(s)$ can be found in appendix \ref{secb}, together with 
more details on our procedure. Here we just mention that, since
the leading contributions are in the real part of 
the running couplings, we replace complex poles
by complex masses and only consider the
real part of the corrections. This in practice means replacing 
\bqa
\mu_{w,z} &\to& M^2_{W,Z}-i\Gamma_{W,Z} M_{W,Z}\,,~~~~
 g^2(\mu_{w,z})~\to~g^2 (M^2_{W,Z})\,,  \nl
c^2_\theta(\mu_z) &\to& c^2_\theta(M^2_Z)\,,~~~~~~~~~~~~~~
T_{W,Z}(\mu_{w,z})~\to~T_{W,Z}(M^2_{W,Z})\,,
\eqa
in the above formulae.
Note that, as for the treatment of the gauge boson widths,
this is equivalent to the fixed width scheme~\cite{passa3}.

When also the $W W \gamma$ coupling contributes, we
introduce, in addition, the following effective three gauge boson vertex 
\begin{center}
\begin{picture}(100,100)(50,-50)
\glin{0,0}{30,0}{4}
\glin{30,0}{60,20}{4}
\glin{30,0}{60,-20}{4}
\Vertex(30,0){2.5}
\LongArrow(5,7)(20,7)
\LongArrow(50,25)(40,17)
\LongArrow(50,-25)(40,-17)
\Text(-3,0)[r]{\small $\gamma_\mu$}
\Text(63,20)[bl]{\small $W^+_\nu$}
\Text(63,-20)[tl]{\small$W^-_\rho$}
\Text(7,12)[bl]{\small $p$}
\Text(44,25)[br]{\small  $p_+$}
\Text(44,-23)[tr]{\small $p_-$}
\Text(80,0)[l]{$= i\,e(s)V_{\mu \nu \rho}$}
\end{picture}
\end{center}
%--
with $s = p^2\,,~~s^+= p^2_+\,,~~s^-= p^2_-$ and
%--
\bqa
V_{\mu \nu \rho} &=& 
 g_{\mu\nu}  (p   -p_+)_\rho
+g_{\nu\rho} (p_+ -p_-)_\mu\,(1+\delta_V)
+g_{\rho\mu} (p_- -p  )_\nu  \nonumber \\
&+&\frac{(p_+ -p_-)_\mu}{s^- - s^+}\left[
 \left(\frac{g(s^-)}{g(s^+)}-1 \right)\,p_{+ \nu}p_{+ \rho}
-\left(\frac{g(s^+)}{g(s^-)}-1 \right)\,p_{- \nu}p_{- \rho}
                             \right] \nonumber \\
\delta_V&=&  \frac{1}{g(s^+) g(s^-) (s^- - s^+)} 
  \left[ g^2(s^+) g^2(s^-)\,[ T_W(s^-)- T_W(s^+) ]  \right. 
\nonumber \\
     &+& \left. [g(s^+)- g(s^-)]\,[ s^- g(s^+) + s^+ g(s^-) ] 
\right]\,. 
\label{modver}
\eqa
It is the easy to see that, with the above choice for $V_{\mu \nu \rho}$,
the $U(1)$ gauge invariance - namely current conservation -
is preserved, even in presence of complex 
masses and running couplings, also with massive final state fermions.

To preserve $U(1)$, one can either
compute $g(s)$ at a fixed scale (for example always at $s= M_W^2$),
while keeping only the running of $e(s)$, or let all the couplings 
run at the proper scale.
 With the first choice the modification of the three gauge boson 
vertex is kept minimal (but the leading running effects are 
still included).
With the second choice everything runs, but a heavier 
modification of the Feynman rules is required.
Since our approach is an effective one, its quality can be 
judged only by comparing with the exact calculation of ref.~\cite{passa1}.
We found that the second choice gives a better agreement
for leptonic single-$W$ final states, while the first one 
is closer to the exact result in the hadronic case, which
is phenomenologically more relevant. Therefore, we adopted this
first option as our default implementation in {\tt NEXTCALIBUR}.

When using {\tt NEXTCALIBUR} in the described {\em running 
coupling mode}, $U(1)$ gauge invariance is preserved but 
$SU(2)$ is, in general, violated. 
The effects of such a violation are small at LEP2 energies. In fact our
method turned out to be numerically equivalent, at $\sqrt{s} \sim 200$ GeV, 
to the {\tt IFL}$_\alpha$ approach of ref.~\cite{ballestrero}.
The authors of ref.~\cite{ballestrero} divide all Feynman diagrams
in gauge invariant sets and use different electromagnetic 
couplings for each set, therefore preserving 
$SU(2) \times U(1)$ gauge invariance.
Our method is more suitable in all cases when no easy separation in
different classes of Feynman diagrams is possible. 

An alternative $SU(2) \times U(1)$ preserving scheme is 
represented by the formalism of ref.~\cite{bbc}, where extra terms
are introduced in the Lagrangian to compensate 
the self-energies contributions and restore gauge invariance.

A version of {\tt NEXTCALIBUR} implementing the equations
of ref.~\cite{bbc} is currently under development.
\section{Structure of the program \label{strupro}}
We shall briefly describe here the general structure of the program.
In the {\tt MAIN} of {\tt NEXTCALIBUR} the input file is read 
and various initializations are performed, for the matrix element evaluation
and the phase space generation.

The phase space is initialized by calling {\tt SUBROUTINE SETPRO},
where the kinematical channels for the phase space generation are build
up, according to the chosen four-fermion final state. 
In {\tt SUBROUTINE SETPRO} also the input parameter set has to be specified. 
The values to be provided are
$G_F$ ({\tt GFERMI}),  
$M_Z$ ({\tt ZM}), 
$M_W$ ({\tt WM}), 
$\Gamma_Z$ ({\tt ZW}), 
$\sin^2 \theta_W$ ({\tt SINW2}), $\alpha_{EM}$ ({\tt ALPHA}), 
and $\Gamma_W$ ({\tt WW}).
 The recommended choice, implemented by default, 
is the $G_F$ scheme described in section \ref{alphar}.
A value for the Higgs mass ({\tt HM}) has to be specified as well and
the corresponding tree level value of $\Gamma_H$ ({\tt HW}) is computed 
with the formula
\bqa
&&\Gamma_H= \frac{\alpha_{EM} M_H}{8 M^2_W \sin^2 \theta_W} 
(m^2_\tau + 3 m^2_b +3 m^2_c) \nonumber \nl
&&m_\tau = 1.777~{\rm GeV},~m_c =  0.75~{\rm GeV},~m_b= 2.9~{\rm GeV}\,.
\eqa
Finally, {\tt SUBROUTINES PHYSICS} and {\tt HELAC\_INIT} are called, 
to initialize the evaluation of the Matrix Element.

The generation of the QED Initial State Radiation is performed
in the {\tt MAIN} of the program. The used algorithm has been 
extensively described in section \ref{sec3}. The number of bins used
in eq.~(\ref{eq9}) to integrate over $t$-dependent Structure Functions 
is set by the variables {\tt nisr10, nisr11, nisr20} and {\tt  nisr21}.
All remaining subroutines are devoted either to the 
matrix element evaluation or to the phase space integration.

Additional cuts, besides those ones specified in the input file, must
be implemented directly in {\tt SUBROUTINE CUTS}, where the two commons

\vskip 0.3cm

{\tt      COMMON/AREA10/PM1(0:4,1:6),PM4(12:65),OMCT1(1:6,3:6)}

{\tt      COMMON/PHOTONS/PM1G(0:4,1:3)}

\vskip 0.3cm

\noindent contain the fermion 
four-momenta computed in the Lab frame ({\tt PM1}),
the invariant mass squared among all possible particle pairs ({\tt PM4}),
the quantities $1 -\cos \theta_{ij}$ ({\tt OMCT1}) and the momenta of the
two ISR photons ({\tt PM1G}).
If the event is rejected ${\tt LNOT}$= 1, and the weight 
is put to zero.

The conventions for the momenta are as follows: 

\begin{tabular}{ll}\\
    {\tt PM1(0,j)} &   is the energy of the j$^{th}$ particle   \\
    {\tt PM1(1,j)} &   is the x component                       \\
    {\tt PM1(2,j)} &   is the y component                       \\
    {\tt PM1(3,j)} &   is the z component                       \\
    {\tt PM1(4,j)} &   is the four-momentum squared.            \\\\
\end{tabular}

Particles number 1 and 2 are the incoming $e^+$ and $e^-$,
respectively. The beam is along the x axis and $e^+$ is 
along the positive direction.

{\tt PM1G(0:4,1)} is the four momentum of the photon 
generated by the incoming $e^+$,
{\tt PM1G(0:4,2)} that one coming from the $e^-$ and              
{\tt PM1G(0:4,3)} is the most energetic between the two.
{\tt PM1G(0:4,j)} is different from zero only when ISR
is generated with a non-vanishing $p_t$ distribution.

The event weight is {\tt W}, the final value of which
is computed immediately before the {\tt FORTRAN} line
\vskip 0.3cm 
{\tt IF (I.GT.0) CALL INBOOK(1,W,IINIT)}.
\vskip 0.3cm 
The status of the run can be checked with the help 
of the file {\tt monitor}, in which the output file name 
and the used Monte Carlo points are printed out
every 5000 iterations (with negative values during the warming phase).
In the file {\tt backup} the intermediate result 
is stored every 50000 Monte Carlo points. 
\section{Compiling instructions}
{\tt NEXTCALIBUR} is written in {\tt FORTRAN} 90. The main program is 
{\tt nextcalibur.f} and there are five included files, namely
\vskip 0.3cm
     {\tt declare.h} 

     {\tt declare\_dp.h} (to run in double precision)

     {\tt declare\_qp.h} (to run in quadruple precision)

     {\tt compl\_mass.h}

     {\tt list.h}.
\vskip 0.3cm
{\tt NEXTCALIBUR} can be run both in double and quadruple precision. 
 Double precision is sufficient to run all processes without electrons
(or positrons) among the final state particles, or processes with final 
state electrons (or positrons) with {\em at most} one final state electron 
(or positron) allowed in the zero-angle forward region.
To run in double precision, typical compiling instructions are:
\begin{verbatim}
  > cp declare_dp.h > declare.h
  > f90 -O nextcalibur.f -o nextcalibur.out
  > nextcalibur.out < nextcalibur.in 
\end{verbatim}
where \verb+nextcalibur.in+ is the input file (see next section). 

To run processes with more than one final state electron (or positron)
allowed in the zero-angle forward region, quadruple precision should
be used, instead. A typical example is the process 
$e^+ e^- \to e^+ e^- \mu^+ \mu^-$ {\em without} any cut.
To run in quadruple precision, typical compiling instructions are:
\begin{verbatim}
  > cp declare_qp.h > declare.h
  > f90 -O -double_size 128 nextcalibur.f -o nextcalibur.out
  > nextcalibur.out < nextcalibur.in 
\end{verbatim}
The flag \verb+-double_size 128+ (that treats double
precision real numbers as quadruple precision real numbers) 
{\em is essential}.
\section{Input \label{input}}
In this section we list the meaning of the input parameters to be specified
by the user:

\vskip 10pt

\noindent {\tt PAR(3) (CHARACTER*2)}

\noindent Produced fermion with label 3 (to be chosen among 
{\tt 'EL','NE','MU','NM',

\noindent 'TA','NT','UQ','DQ','CQ','SQ','TQ','BQ'}).\\

\noindent {\tt PAR(4) (CHARACTER*2)}

\noindent Produced fermion with label 4.\\ 

\noindent {\tt PAR(5) (CHARACTER*2)}

\noindent Produced fermion with label 5.\\ 

\noindent {\tt PAR(6) (CHARACTER*2)}

\noindent Produced fermion with label 6.\\

\noindent {\tt NIPT (INTEGER)}

\noindent The number of points for the Monte Carlo integration.\\         

\noindent {\tt NWARM (INTEGER)}

\noindent Number of points for the warming up of the
phase space generation but not used for the actual computation. 
For processes with zero angle electrons in the final state
it is recommended to chose a non zero value (e.g. {\tt NWARM}=10000).\\     

\noindent {\tt NOPT (INTEGER)}

\noindent Number of points for the phase space optimization
but also used for the computation.\\         

\noindent {\tt ISPEPMAX (INTEGER)}

\noindent Number of iterations for optimizing the a-priori weights.\\

\noindent {\tt OUTPUTNAME (CHARACTER*15)}

\noindent The name of the output file.\\

\noindent {\tt KREL (INTEGER)}

\noindent Selects the kinematical channels to be used for 
the phase space generation. {\tt KREL=0} is the recommended value
for normal runs.\\

\noindent {\tt LQED (INTEGER)}

\noindent It includes (1) or excludes (0) ISR.\\

\noindent {\tt LQ\_PT1 (INTEGER)}

\noindent Collinear QED radiation from the incoming $e^+$ (0) or
          generation of ISR photons with non zero $p_t$ (1).\\

\noindent {\tt LQ\_PT2 (INTEGER)}

\noindent Collinear QED radiation from the incoming $e^-$ (0) or
          generation of ISR photons with non zero $p_t$ (1).\\

\noindent {\tt LQ\_SC1 (INTEGER)}

\noindent  $t$-channel ISR scale for the incoming 
           $e^+$ off (0) or on (1). When choosing (0), ISR is computed  
           with the scale $Q^2= s$. When choosing (1), if   
           the process has at least 1 final               
           state $e^+$, the used scale is $Q^2= |t|$.          
           When there are 2 $e^+$ in the final state,        
           the used scale is $Q^2= min(|t|)$.\\               

\noindent {\tt LQ\_SC2 (INTEGER)}

\noindent  $t$-channel ISR scale for the incoming 
           $e^-$ as above.\\

\noindent {\tt IWIDTH (INTEGER)}

\noindent Option for the treatment of boson masses and 
          running of $\alpha_{EM}$. If {\tt IWIDTH=1}      
          all boson masses are taken to be complex,
          also in the couplings.            
          When {\tt IWIDTH=2} $\alpha_{EM}$ is running and the        
          Modified Fermion-Loop is switched on.\\           

\noindent {\tt IHIGGS (INTEGER)}

\noindent  Higgs diagrams included (1) or not (0).\\

\noindent {\tt REL (REAL*8)}

\noindent  Electron mass.\\

\noindent {\tt RNE (REAL*8)}

\noindent  Mass of $\nu_e$.\\

\noindent {\tt RMU (REAL*8)}

\noindent  Muon mass.\\

\noindent {\tt RNM (REAL*8)}

\noindent  Mass of $\nu_{\mu}$.\\

\noindent {\tt RTA (REAL*8)}

\noindent  Tau mass.\\

\noindent {\tt RNT (REAL*8)}

\noindent  Mass of $\nu_{\tau}$.\\

\noindent {\tt RUQ (REAL*8)}

\noindent  Mass of the $u$ quark.\\

\noindent {\tt RDQ (REAL*8)}

\noindent  Mass of the $d$ quark.\\

\noindent {\tt RCQ (REAL*8)}

\noindent  Mass of the $c$ quark.\\

\noindent {\tt RSQ (REAL*8)}

\noindent  Mass of the $s$ quark.\\

\noindent {\tt RTQ (REAL*8)}

\noindent  Mass of the $t$ quark.\\

\noindent {\tt RBQ (REAL*8)}

\noindent  Mass of the $b$ quark.\\

\noindent {\tt ROOTS (REAL*8)}

\noindent The total energy of the colliding $e^+$ and $e^-$. All energies
are in GeV.\\

\noindent {\tt SHCUT (REAL*8)}

\noindent Minimum value of the invariant mass squared of the event 
after ISR.\\

\noindent {\tt ECUT(3) (REAL*8)}

\noindent Minimum energy of particle number 3.\\

\noindent {\tt ECUT(4) (REAL*8)}

\noindent Minimum energy of particle number 4.\\

\noindent {\tt ECUT(5) (REAL*8)}

\noindent Minimum energy of particle number 5.\\

\noindent {\tt ECUT(6) (REAL*8)}

\noindent Minimum energy of particle number 6.\\

\noindent {\tt SCUT(3,4) (REAL*8)}

\noindent Minimum value of $(\,p(3)\,+\,p(4)\,)^2$. All invariant masses 
in GeV$^2$.\\

\noindent {\tt SCUT(3,5) (REAL*8)}

\noindent Minimum value of $(\,p(3)\,+\,p(5)\,)^2$.\\

\noindent {\tt SCUT(3,6) (REAL*8)}

\noindent Minimum value of $(\,p(3)\,+\,p(6)\,)^2$.\\

\noindent {\tt SCUT(4,5) (REAL*8)}

\noindent Minimum value of $(\,p(4)\,+\,p(5)\,)^2$.\\

\noindent {\tt SCUT(4,6) (REAL*8)}

\noindent Minimum value of $(\,p(4)\,+\,p(6)\,)^2$.\\

\noindent {\tt SCUT(5,6) (REAL*8)}

\noindent Minimum value of $(\,p(5)\,+\,p(6)\,)^2$.\\

\noindent {\tt CMAX(1,3) (REAL*8)}

\noindent Maximum value of $\cos \theta$ between particle 1 and 3.\\

\noindent {\tt CMAX(1,4) (REAL*8)}

\noindent Maximum value of $\cos \theta$ between particle 1 and 4.\\

\noindent {\tt CMAX(1,5) (REAL*8)}

\noindent Maximum value of $\cos \theta$ between particle 1 and 5.\\

\noindent {\tt CMAX(1,6) (REAL*8)}

\noindent Maximum value of $\cos \theta$ between particle 1 and 6.\\

\noindent {\tt CMAX(2,3) (REAL*8)}

\noindent Maximum value of $\cos \theta$ between particle 2 and 3.\\

\noindent {\tt CMAX(2,4) (REAL*8)}

\noindent Maximum value of $\cos \theta$ between particle 2 and 4.\\

\noindent {\tt CMAX(2,5) (REAL*8)}

\noindent Maximum value of $\cos \theta$ between particle 2 and 5.\\

\noindent {\tt CMAX(2,6) (REAL*8)}

\noindent Maximum value of $\cos \theta$ between particle 2 and 6.\\

\noindent {\tt CMAX(3,4) (REAL*8)}

\noindent Maximum value of $\cos \theta$ between particle 3 and 4.\\

\noindent {\tt CMAX(3,5) (REAL*8)}

\noindent Maximum value of $\cos \theta$ between particle 3 and 5.\\

\noindent {\tt CMAX(3,6) (REAL*8)}

\noindent Maximum value of $\cos \theta$ between particle 3 and 6.\\

\noindent {\tt CMAX(4,5) (REAL*8)}

\noindent Maximum value of $\cos \theta$ between particle 4 and 5.\\

\noindent {\tt CMAX(4,6) (REAL*8)}

\noindent Maximum value of $\cos \theta$ between particle 4 and 6.\\

\noindent {\tt CMAX(5,6) (REAL*8)}

\noindent Maximum value of $\cos \theta$ between particle 5 and 6.\\

A last remark is in order.
When running with the options {\tt LQ\_PT1= 1} and {\tt LQ\_PT2= 1} for
processes with more than one
 $e^-$ (or $e^+$) in the final state, even by cutting out 
the beam cone the generation efficiency of {\tt NEXTCALIBUR} may 
be bad, due to genuine $\gamma \gamma$ like events {\em kicked out} 
from the beam cone by large $p_t$ photons.
If one is interested just in the total cross section, running
with {\tt LQ\_PT1 = LQ\_PT2= 0} solves the problem, 
otherwise a reasonable cut on the $p_t$ of the generated photons must
be applied to restore the efficiency.
\section{Test Run Output \label{tro}}
We conclude our description with an example of a calculation that
can be performed with {\tt NEXTCALIBUR}. Notice that no cut 
is present on the final state particles.
One should be able to reproduce
this output within the estimated Monte Carlo error.
Using an input file as follows
\clearpage
\begin{verbatim}
el               ! par(3)       - produced fermion                   
ne               ! par(4)       - produced antifermion                       
uq               ! par(5)       - produced fermion                           
dq               ! par(6)       - produced antifermion                       
250000           ! nipt         - number of Monte Carlo points               
10000            ! nwarm        - points for the warming
40000            ! nopt         - points for a.p.weights optim
2                ! istepmax     - iterations for a.p.weights optim            
output           ! outputname   - program name                               
0                ! krel   (0:5) - Channels for MC Mapping                    
1                ! lqed   (0,1) - QED off or on                              
0                ! lq_pt1 (0,1) - finite pt for incoming e(+)                
0                ! lq_pt2 (0,1) - finite pt for incoming e(-)                
1                ! lq_sc1 (0,1) - t-chan scale for incoming e(+)             
1                ! lq_sc2 (0,1) - t-chan scale for incoming e(-)             
2                ! iwidth (1:2) - option for boson masses and running        
0                ! ihiggs (0:1) - Higgs included or not                      
0.51099906d-3    ! rel          - fermion masses:                            
0.d0             ! rne          -                                            
105.658389d-3    ! rmu          -                                            
0.d0             ! rnm          -                                            
1.77705d0        ! rta          -                                            
0.d0             ! rnt          - be aware that when                         
0.005d0          ! ruq          - fermion masses are zero                    
0.010d0          ! rdq          - the computation gets faster!               
1.55d0           ! rcq          -                                            
150.d-3          ! rsq          -                                            
175.d0           ! rtq          -                                            
4.7d0            ! rbq          -                                            
200.d0           ! roots        - center of mass energy (GeV)                
0.d0             ! shcut        - cut on inv. mass squared after QED         
0.d0             ! ecut_3       - Energy cut for particle 3                  
0.d0             ! ecut_4       - Energy cut for particle 4                  
0.d0             ! ecut_5       - Energy cut for particle 5                  
0.d0             ! ecut_6       - Energy cut for particle 6                  
0.d0             ! scut_34      - inv. mass cut for the system (34)          
0.d0             ! scut_35      - inv. mass cut for the system (35)          
0.d0             ! scut_36      - inv. mass cut for the system (36)          
0.d0             ! scut_45      - inv. mass cut for the system (45)          
0.d0             ! scut_46      - inv. mass cut for the system (46)          
0.d0             ! scut_56      - inv. mass cut for the system (56)          
1.d0             ! cmax_13      - cut on cos_(1,3)                           
1.d0             ! cmax_14      - cut on cos_(1,4)                           
1.d0             ! cmax_15      - cut on cos_(1,5)                           
1.d0             ! cmax_16      - cut on cos_(1,6)                           
1.d0             ! cmax_23      - cut on cos_(2,3)                           
1.d0             ! cmax_24      - cut on cos_(2,4)                           
1.d0             ! cmax_25      - cut on cos_(2,5)                           
1.d0             ! cmax_26      - cut on cos_(2,6)                           
1.d0             ! cmax_34      - cut on cos_(3,4)                           
1.d0             ! cmax_35      - cut on cos_(3,5)                           
1.d0             ! cmax_36      - cut on cos_(3,6)                           
1.d0             ! cmax_45      - cut on cos_(4,5)                           
1.d0             ! cmax_46      - cut on cos_(4,6)                           
1.d0             ! cmax_56      - cut on cos_(5,6)                           
\end{verbatim}
and the input values

\begin{tabular}{llllll}\\
$G_{F}   $&=&$ 1.6637~10^{-5}$ ${\rm GeV}^{-2}$  &
$M_Z     $&=&$ 91.1867$ GeV     \\ 
$M_W     $&=&$ 80.35  $ GeV     &
$\Gamma_Z$&=&$ 2.4939  $ GeV    \\ 
$\Gamma_W$&=&$ \frac{3\,G_F\,M^2_W}{\pi\sqrt{8}} $ & 
$\sin^2 \theta_W$&=&$ 1-\frac{M^2_W}{M^2_Z}$\\ 
$\alpha  $&=&$ \sin^2_W\,G_F\,M_W^2\,\frac{\sqrt{2}}{\pi}$\,,&&&
\end{tabular}

\vspace{0.5cm}
\noindent we get the following output file
\begin{verbatim}
                    output         


  process :    antiel(1) el(2) ---> el(3) antine(4) uq(5) antidq(6)

  This run is with:
                   
  nipt    =     250000
  nwarm   =      10000
  nopt    =      40000
  istepmax=          2
  krel    =          0
  lqed    =          1
  lq_pt1  =          0
  lq_pt2  =          0
  lq_sc1  =          1
  lq_sc2  =          1
  iwidth  =          2
  ihiggs  =          0
  rel     =    0.510999D-03
  rne     =    0.000000D+00
  rmu     =    0.105658D+00
  rnm     =    0.000000D+00
  rta     =    0.177705D+01
  rnt     =    0.000000D+00
  ruq     =    0.500000D-02
  rdq     =    0.100000D-01
  rcq     =    0.155000D+01
  rsq     =    0.150000D+00
  rtq     =    0.175000D+03
  rbq     =    0.470000D+01
  roots   =    0.200000D+03
  shcut   =    0.000000D+00
  ecut_3  =    0.000000D+00
  ecut_4  =    0.000000D+00
  ecut_5  =    0.000000D+00
  ecut_6  =    0.000000D+00
  scut_34 =    0.000000D+00
  scut_35 =    0.000000D+00
  scut_36 =    0.000000D+00
  scut_45 =    0.000000D+00
  scut_46 =    0.000000D+00
  scut_56 =    0.000000D+00
  cmax_13 =    0.100000D+01
  cmax_14 =    0.100000D+01
  cmax_15 =    0.100000D+01
  cmax_16 =    0.100000D+01
  cmax_23 =    0.100000D+01
  cmax_24 =    0.100000D+01
  cmax_25 =    0.100000D+01
  cmax_26 =    0.100000D+01
  cmax_34 =    0.100000D+01
  cmax_35 =    0.100000D+01
  cmax_36 =    0.100000D+01
  cmax_45 =    0.100000D+01
  cmax_46 =    0.100000D+01
  cmax_56 =    0.100000D+01

  Input parameters: 

  gfermi  =    0.116637D-04
  zm      =    0.911867D+02
  wm      =    0.803500D+02
  zw      =    0.249390D+01
  ww      =    0.204277D+01
  sin2w   =    0.223558D+00
  1/alpha =    0.131958D+03
  hm      =    0.120000D+03
  hw      =    0.236865D-02

  --------------------------------------------
                                       
  Kinematical Diagrams:                
                                       
      channel                     permutation
                                       
  1:  annihi2(wm)                 1 2 3 4 5 6
  2:  annihi2(wm)                 1 2 4 3 5 6
  3:  annihi2(wm)                 1 2 5 6 3 4
  4:  annihi2(wm)                 1 2 6 5 3 4
  5:  bremf4(wm)                  1 2 3 4 5 6
  6:  bremb2(wm)                  2 1 3 4 5 6
  7:  conver3(wm,wm)              1 2 5 6 3 4
  8:  multi2                      2 1 3 5 6 4
  9:  multi3                      1 2 3 5 6 4
 10:  nonab1(wm)                  1 2 3 4 5 6
 11:  nonab2(wm)                  2 1 3 5 6 4
 12:  nonab4(wm)                  1 2 3 5 6 4
 13:  rambo4                      1 2 3 4 5 6
                    
  --------------------------------------------
                    
  Differences in the computation 
  of the a-priori weights:       
                    
  diff(           1 )=    5.38443233306179     
  diff(           2 )=    6.18288649090395     
  diff(           3 )=    2.25190931218116     
                    
  a-priori weights: 
                    
  1 :     0.301826D-01
  2 :     0.202852D-01
  3 :     0.210927D-01
  4 :     0.327752D-01
  5 :     0.251601D-01
  6 :     0.873505D-01
  7 :     0.299054D+00
  8 :     0.185762D+00
  9 :     0.199223D-02
 10 :     0.162069D+00
 11 :     0.104346D+00
 12 :     0.251601D-01
 13 :     0.477138D-02
                    
  --------------------------------------------

  Cross Section (pb):            
                                 
   1  0.79733D+00  +/-  0.46D-02

\end{verbatim}
\appendix
\section{The kinematical channel {\tt MULTI1} \label{seca}}
We refer to fig. 1 and split the integration over 
the massive four-body phase space into a 3-body decay 
times a 2-body decay
\bqa
\int d(PS)_4 = \int ds_{45}\,I_3(p_3,p_6,p_{45})\,I_2(p_4,p_5)\,,
\eqa
with
\bqa
&&\!\!\!I_3(p_3,p_6,p_{45}) = 
\int D(p_3,m^2_3)\,D(p_6,m^2_6)\,D(p_{45},s_{45})\,
\delta^4(p_{12}-p_3-p_6-p_{45})\,,\nl
&&\!\!\!I_2(p_4,p_5) = \int D(p_4,m^2_4)\,D(p_5, m^2_5)\, 
\delta^4(p_{45}-p_4-p_5)\,, \nl
&&\!\!\!D(p_i,m^2_i)= d^4p_i\,\delta(p^2_i-m^2_i)\,\theta(E_i)\,,~~
p_{ij}= p_i+p_j\,,~~s_{45}= p^2_{45}\,.
\eqa
The three-body phase space can be rewritten as
\bqa
&&I_3(p_3,p_6,p_{45})= \frac{s}{32}\int_{2\sqrt{\mu_3}}^{x_-} dx \int dy\,\,
d\Omega_3\,d\Omega_6\,\delta[F(x,y)]\,,
\label{eq11}
\eqa
where $x$ and $y$ are the reduced
energies of the particles 3 and 6
$$E_3= \frac{\sqrt{s}}{2}x\,,~~E_6= \frac{\sqrt{s}}{2}y\,,$$
$d \Omega_i= d\phi_i\,d\cos \theta_i$ is 
the integration element over the solid angle of particle $i$ 
and $c_{36}$ the cosine of the angle between particles 3 and 6.
Finally,
\bqa
&& F(x,y)= 2\,G(x,y)/\sqrt{H(x,y)}-c_{36}\,,~~
   G(x,y)= M-x-y+\frac{xy}{2}\,,\nl
&& H(x,y)= (x^2-4\mu_3)(y^2-4\mu_6)\,,~~
   M = 1+\mu_3+\mu_6-\mu\,,\nl
&& \mu_i= \frac{m^2_i}{s}\,,~~\mu= \frac{s_{45}}{s}\,,~~
 x_-= (1+\mu_3) - (\sqrt{\mu}+\sqrt{\mu_6})^2\,.
\label{eq12}
\eqa
Eq. (\ref{eq11}) is equivalent to
\bqa
&&I_3(p_3,p_6,p_{45})= \nl
&&\frac{s}{32}\int_{2\sqrt{\mu_3}}^{x_-} dx \int dy\,\,
d\Omega_3\,d\Omega_6\,\frac{1}{|F^\prime(x,y)|}
\left[
\delta(y-y_+) + \delta(y-y_-)
\right]\,,
\label{eq13}
\eqa
where
\bqa
&&F^\prime(x,y)= \frac{\partial F(x,y)}{\partial y}\,,~~~
  y^\pm =\frac{\beta \pm \sqrt{\Delta}}{\alpha}\,,
~~~\Delta= \beta^2-\alpha \gamma\,,\nl
&& \alpha= 4\left(1-\frac{x}{2} \right)^2 -c^2_{36}\,(x^2-4\mu_3) \,,~~~
   \beta =  4\left(1-\frac{x}{2} \right) (M-x)\,,\nl
&& \gamma= 4(M-x)^2+4c^2_{36}\,\mu_6 (x^2-4\mu_3)\,.  
\eqa
As for the two-body decay, the following expression holds
\bqa
I_2(p_4,p_5) &=& \frac{1}{8 s_{45}} \lambda({s_{45},m^2_4,m^2_5})
\int d\Omega_4\,,\nl
\lambda(a,b,c)&=& \sqrt{a^2+b^2+c^2-2ab -2ac -2bc}\,.
\eqa

The first four steps of the generation algorithm are as follows:
\begin{itemize}
\item[1)] generate $s_{45}$ with distribution $$\frac{1}{s_{45}}\,;$$
\item[2)] generate $x$ with distribution $$\frac{1}{1-x}\,;$$  
\item[3)] generate $\phi_3$ uniformly and $c_3 \equiv \cos \theta_3$ with
          distribution 
          $$\frac{1}{a_3-c_3}\,,~~~a_3 = 
           \frac{E_1 E_3-m^2_e}{|\vec{p}_1||\vec{p}_3|}\,;$$
\item[4)] generate $\phi_6$ uniformly and $c_6 \equiv \cos \theta_6$ with
          distribution 
          \bqa 
          \frac{1}{a_6-c_6}\,,~~~a_6 = 
           \frac{E_2 \tilde{E}_6-m^2_e}{|\vec{p}_2||\vec{\tilde p}_6|}\,,~~~
          \tilde{E}_6= \frac{\sqrt{s}}{2}\tilde y\,,~~~
          |\vec{\tilde p}_6|= \sqrt{\tilde{E}^2_6-m^2_e}\,, 
          \eqa
          where $\tilde y$ is the solution of the equation
          \bqa
          2\,G(x,\tilde y)/\sqrt{H(x,\tilde y)}+1 = 0\,.
          \eqa
\end{itemize}
The reason for the above choice of $a_6$ is that,
at this stage of generation, $E_6$ is unknown, but its value
when $c_{36}= -1$ (namely when the multi-peripheral singularity
is more pronounced) can be computed using the condition
$F(x,\tilde y)= 0$, as read from eq. (\ref{eq12}) with $c_{36}= -1$.

 From the generated values of $c_3$, $c_6$, $\phi_3$ and $\phi_6$
one then computes 
$$c_{36}= \cos \theta_3 \cos \theta_6 + 
          \sin \theta_3 \sin  \theta_6 \cos (\phi_3-\phi_6)\,,$$ 
to be compared with two possible values ($c_{36}^\pm$) 
obtained from the conditions $y= y^{\pm}$ in eq. (\ref{eq13}).
 There are four possibilities. 
If $c_{36} \ne c_{36}^\pm$ the event has vanishing weight.
If $c_{36} = c_{36}^+ \ne c_{36}^-$, then
$y= y^+$ and only the first term in eq. (\ref{eq13}) contributes.
If $c_{36} = c_{36}^- \ne c_{36}^+$, then 
the second term in eq. (\ref{eq13}) has to be chosen.
Finally, if $c_{36} = c_{36}^- = c_{36}^+$, the two terms
are picked up randomly and the total event weight multiplied by 2.

The last part of the generation  accounts for the $t$-channel
exchanged fermion in {\tt MULTI1}.
By defining $q= p_1-p_3$, the variable $c_4 \equiv \cos \theta_4$
peaks, in the rest frame of $s_{45}$, as follows
\bqa
\frac{1}{a-c_4}\,,
~~~a= \frac{2 q_0 E_4-q^2}{2 |\vec{q}| |\vec{p}_4|}\,,~~~
E_4= \frac{s_{45}+m^2_4-m^2_5}{2 \sqrt{s_{45}}}\,.
\eqa
On the other hand, also the diagram obtained when replacing  
$p_4 \leftrightarrow p_5$ contributes with an analogous
peaking behaviour
\bqa
\frac{1}{b-c_5}\,,
~~~b= \frac{2 q_0 E_5-q^2}{2 |\vec{q}| |\vec{p}_5|}\,,~~~
E_5= \frac{s_{45}+m^2_5-m^2_4}{2 \sqrt{s_{45}}}\,.
\eqa
Since, in the rest frame of $s_{45}$, $c_4= -c_5$,
both distributions can be simultaneously mapped, 
via the last three items of our generation algorithm:
\begin{itemize}
\item[5)] go the rest frame of $s_{45}$;
\item[6)] generate $\phi_4$ uniformly and $c_4$ with distribution 
          $$\frac{1}{(a-c_4)(b+c_4)}\,;$$
\item[7)] boost $p_4$ and $p_5$ back to the lab frame.
\end{itemize}
A massive four-fermion event is then generated, taking
into account all peaks due to $t$-channel exchanged 
massless or nearly massless particles.

 In {\tt NEXTCALIBUR} the described procedure is 
further refined by making it fully symmetric under the replacements
$\{p_1,p_3\} \leftrightarrow \{p_2,p_6\}$.

 Finally, the same ingredients are used also when constructing 
the channel {\tt MULTI2}, the only variation being a different choice 
of the distributions employed in the generation of some of
the variables.
\section{The Modified Fermion-Loop approach \label{secb}}
Following ref.~\cite{passa2} we split the bare W and Z boson 
self-energies into universal and non universal contributions 
\bqa
\hat \Sigma_W(p^2) &=& \hat {\bar \Sigma}_W(p^2)+ 
{\hat g}^2 T_W(p^2)\,, \nl 
\hat \Sigma_Z(p^2) &=& \hat {\bar \Sigma}_Z(p^2)+ 
\frac{{\hat g}^2}{{\hat c}^2} T_Z(p^2)\,, 
\eqa
where ${\hat g}$ is the bare weak coupling constant 
${\hat s}$ the sine of the bare mixing angle and  
${\hat c}^2 \equiv 1 - {\hat s}^2$.
When including fermion loops only, the explicit 
expressions for the above quantities are
\bqa
\hat {\bar \Sigma}_W(p^2)&=& \frac{{\hat g}^2}{48 \pi^2} \frac{1}{2}
\sum_f N^c_f \left  \{
I(p^2)+(1+2 I_{3f} Y_f) F(p^2,m_f)
             \right \}\,, \\
\hat {\bar \Sigma}_Z(p^2)&=& \frac{{\hat g}^2}{48 {\hat c}^2 \pi^2} 
\sum_f N^c_f \left  \{
({\hat v}^2_f+{\hat a}^2_f) I(p^2) \right. \nl
&+& \left. ({\hat v}^2_f+{\hat a}^2_f+4 I_{3f} Y_f {\hat a}^2_f) F(p^2,m_f)
             \right \}\nonumber \,,
\eqa
with $Y_f= 2 (Q_f-I_{3f})$, $v_f= I_{3f}-2 {\hat s}^2 Q_f$
, $a_f= -I_{3f}$ and $N^c_f=$ number of colors. Furthermore
\bqa
I(p^2)   &=& -p^2 f(\epsilon) \left[ 
    \frac{2}{\epsilon} + \ln(-p^2)-\frac{5}{3}
                         \right]\,,\nl
F(p^2,m) &=& -6 p^2 f(\epsilon)\,G(p^2,p^2,m)\,,\nl
G(s,t,m) &=&\int_0^1 dx \left\{x(1-x) \ln \frac{m^2-s x (1-x)}{-t x (1-x)}
                                \right\}\,, \nl
f(\epsilon) &=& \pi^\frac{\epsilon}{2} \Gamma \left( 1-\frac{\epsilon}{2} 
\right)\,, ~~~~{\rm with}~~\epsilon = n_d-4\,.
\eqa
Finally, the leading contribution to $T_{W,Z}(p^2)$ originate from
the top quark:
\bqa
T_{W}(p^2) &=& f(\epsilon)\frac{3}{16 \pi^2} \left\{
 m^2_t \left[\frac{1}{\epsilon} + \frac{1}{2} \ln m^2_t - \frac{1}{4}+ J_0(b)
-J_1(b) \right] \right. \nl
&-& \left. 2 p^2 \left[
J_1(b)-J_2(b)+\frac{5}{36}-\frac{1}{6} \ln(b) - \frac{2}{3} G(p^2,p^2,m_t) \right] \right\}\,, \nl
T_{Z}(p^2) &=& f(\epsilon)\frac{1}{16 \pi^2} \left\{
\frac{3}{2}m^2_t \left[ \frac{2}{\epsilon} + \ln m^2_t + J_0(b_+) + J_0(b_-)
         \right] \right. \nl
&+& \left. p^2\,G(p^2,p^2,m_t) \right \}\,,
\eqa
where
\bqa
J_i(b) = \int_0^1 dx\,\, x^i\,\ln(1+bx)\,,~b = -\frac{p^2}{m^2_t}\,,
~b_\pm= \frac{1}{x_\pm}\,,~x_\pm^2-x_\pm+\frac{1}{b}= 0\,.
\eqa
Two more expressions are necessary, namely 
the photon (${\hat \Sigma}_\gamma$) and  the photon-Z mixing 
(${\hat \Sigma}_X$) self-energies
\bqa
{\hat \Sigma}_\gamma (p^2) &=& \frac{{\hat g}^2 {\hat s}^2}{12 \pi^2}
\sum_f Q^2_f N^c_f \left  \{
I(p^2)+F(p^2,m_f)
             \right \}\,, \\
{\hat \Sigma}_X(p^2)  &=&-\frac{{\hat g}^2 }{24 \pi^2} 
\frac{{\hat s}}{{\hat c}}
\sum_f {\hat v}_f Q_f N^c_f \left  \{
I(p^2)+F(p^2,m_f)
             \right \}\,.
\eqa
In terms of self-energies the running couplings are
defined as follows
\bqa
\frac{e^2(p^2)}{{\hat g}^2{\hat s}^2} = \frac{p^2}{p^2+
\hat \Sigma_\gamma(p^2)}\,,~
\frac{g^2(p^2)}{{\hat g}^2     } = \frac{p^2}{p^2+\hat 
{\bar \Sigma}_W(p^2)}\,,~
s^2(p^2) = \frac{e^2(p^2)}{g^2(p^2)}\,.
\eqa
Since, as explained in section~\ref{alphar}, 
we are only interested in the real
part of the corrections, we renormalize the W and Z boson 
propagators at the real mass 
\bqa
M^2_W &=& \hat \mu_w- {\rm Re}\,\hat \Sigma_W(M^2_W)\,,\nonumber \\
M^2_Z &=& \hat \mu_z- {\rm Re}\,\hat Z(M^2_W)\,,
\eqa
where $\hat \mu_{w,z}$ are the bare gauge boson masses squared and
\bqa
\hat Z(p^2)= \hat \Sigma_Z(p^2)- 
\frac{{\hat \Sigma}^2_X(p^2)}{p^2+\hat \Sigma_\gamma(p^2)}\,.
\eqa
Finally, $G_F$ fixes our third fitting equation
\bqa
\frac{G_F}{\sqrt{2}} =  \frac{{\hat g}^2}{8\,
(\hat \mu_W- \hat \Sigma_W(0))}\,.
\eqa
By solving the above equations it is easy to get $\hat g^2$, ${\hat \mu}_w$
and ${\hat s}^2$  in terms of $G_F$ $M^2_W$ and $M^2_Z$.
Putting everything together, the running couplings read
\bqa
g^2(s) &=& M^2_W \,a \,{\rm Re}\,\left\{
1+\frac{3}{2} M^2_W \frac{a}{\pi^2}\,G(M^2_W,s,0)
- \frac{M^2_W a}{8 \pi^2} \left[ G(s,s,m_e) \nonumber \right. \right.
\nonumber \\
&+&  G(s,s,m_\mu)+G(s,s,m_\tau)
+2\,G(s,s,m_u) +2\,G(s,s,m_c)\nonumber \\
&+&  2\,G(s,s,m_t)-2\,G(M^2_W,M^2_W,m_t)
+G(s,s,m_d) \nonumber \\
&+& \left. \left. G(s,s,m_s)+G(s,s,m_b)\right] -a\, 
\left[T_W(M^2_W)-T_W(0)\right] 
\right\}^{-1}\,,
\label{grun}
\eqa
and
\bqa
e^2(s) &=& M^2_W a N\,{\rm Re}\, \left\{ \left[
1-a [T_W(M^2_W)-T_W(0)]+\frac{3}{2} \frac{M^2_W a}{\pi^2}
G(M^2_W,M^2_Z,0) \right. \right. \nonumber \\
&+& \left.\frac{1}{4} \frac{M^2_W a}{\pi^2} \left(
G(M^2_W,M^2_W,m_t)-G(M^2_Z,M^2_Z,m_t) \right) \right]^2 \nonumber \\
&-& \frac{4}{\pi^2} M^2_W a N \left[ G(s,M^2_Z,0) + \frac{1}{8} 
[G(s,s,m_e)+G(s,s,m_\mu) 
\right. \nonumber \\
&+& G(s,s,m_\tau)] + \frac{1}{6} [G(s,s,m_u)+G(s,s,m_c)+G(s,s,m_t)]
\nonumber \\
&-& \frac{1}{6} G(M^2_Z,M^2_Z,m_t) 
+\frac{1}{24}[G(s,s,m_d)+G(s,s,m_s)] \nonumber \\
&+& \left.\left. \frac{1}{24} G(s,s,m_b) \right] \right\}^{-1}\,,
\label{erun}
\eqa
where
\bqa
N &=& 1-\frac{M^2_W}{M^2_Z} \left[1-a\,{\rm Re}(T_Z(M^2_Z)-T_W(0)) \right]
  \nonumber \\
  &-& a\,{\rm Re}\,(T_W(M^2_W)-T_W(0)) +\frac{3}{2}\frac{M^2_W a}{\pi^2}
G(M^2_W,M^2_Z,0) 
  \nonumber \\
  &+& \frac{1}{4} \frac{M^2_W a}{\pi^2} \,
      {\rm Re}\,\left[G(M^2_W,M^2_W,m_t)-G(M^2_Z,M^2_Z,m_t) \right]\,,
  \nonumber \\
a &=& \frac{8 G_F}{\sqrt{2}}\,.
\eqa
A comment is in order on the appearance of the light quark masses in 
the above expressions.
The weak coupling $g^2(s)$ gives sizable contributions 
at high energies, where all light fermion masses can be neglected, 
so that only the top terms survive in eq.~(\ref{grun}).
On the contrary, being interested in the low energy regime
of eq. ~(\ref{erun}), we have to include in our formulae
the hadron vacuum polarization $\Delta^h_r(s)$, 
as computed in ref.~\cite{jeger}, rather than using quark masses.
By inserting
\bqa
{\rm Re}\, G(s,s,m) &=& \frac{1}{6} \ln \frac{m^2}{|s|}+\frac{5}{18}
+ {\rm Re}\,J(s,m)\,, \nonumber \\
J(s,m) &=& \int_0^1 dx\,x(1-x)\ln \left(1-\frac{s}{m^2} x(1-x)\right)
\eqa
in eq. ~(\ref{erun}), one gets
\bqa
e^2(s) &=& M^2_W a N\,{\rm Re}\, \left\{ \left[
1-a [T_W(M^2_W)-T_W(0)]+\frac{3}{2} \frac{M^2_W a}{\pi^2}
G(M^2_W,M^2_Z,0) \right. \right. \nonumber \\
&+& \left.\frac{1}{4} \frac{M^2_W a}{\pi^2} \left(
G(M^2_W,M^2_W,m_t)-G(M^2_Z,M^2_Z,m_t) \right) \right]^2 \nonumber \\
&+& \frac{2}{3 \pi^2} M^2_W a N \left[ J(M^2_Z,m_t) -
 J(s,m_t) \right]  \nonumber \\
&-& \frac{4}{\pi^2} M^2_W a N \left[
\frac{25}{108} + \frac{1}{48} \sum_{\ell= e, \mu, \tau} 
\ln \frac{m^2_\ell}{M^2_Z}
+\frac{1}{36}  \sum_{q= u,c} \ln \frac{m^2_q}{M^2_Z}
\right.
\nonumber \\
&+& \left. \left. \frac{1}{144}  \sum_{q= d,s,b} \ln \frac{m^2_q}{M^2_Z}
+\frac{1}{8}  \sum_{\ell= e,\mu,\tau} J(s,m_\ell)
+ \frac{\pi}{16 \alpha(0)} \Delta^h_r(s)
\right] \right\}^{-1}\,,
\label{erun1}
\eqa
with $1/\alpha(0)=137.0359895$.
Notice that, due to our choice of the input parameter set, 
a dependence on the light quark masses is left in the previous equation. 
However, since the pure bosonic corrections are missing, $m_t$ 
can be interpreted as a free parameter~\cite{passa2} to be fitted
in such a way that $\alpha(0) \equiv e^2(0)/4 \pi= 1/137.0359895$.
The quark masses appearing in eq.~(\ref{erun1}) only 
affect this fitting procedure, and not the running 
of the electro-magnetic
coupling. In other words, a change in the $m_q$'s is compensated
by a different fitted value for $m_t$.
 For consistency, in {\tt NEXTCALIBUR} the light quark masses are taken 
to be those that reproduce the hadronic vacuum polarization, 
but the numerical results are rather insensitive to the
initial choice of $m_q$'s.

Eqs.~(\ref{grun}) and~(\ref{erun1}), together with equations 
(\ref{bosprop})-(\ref{modver}) completely define our 
Modified Fermion-Loop approach.
\section*{Acknowledgements}
Discussions with the participants in the CERN LEP2 Monte
Carlo Workshop~\cite{wshop} are acknowledged.
The authors also acknowledge the financial support of the European 
Union under contract HPRN-CT-2000-00149.


\begin{thebibliography}{999}
\bibitem{lep2} D.~Bardin et al., in {\em Physics at LEP2}, CERN 96-01 (1996),
 eds. G.~Altarelli, T.~Sj\"ostrand and F.~Zwirner, Vol 2, p. 3.
\bibitem{wshop} M.~W.~Gr\"unewald et al., in
 {\em Reports of the working groups on precision calculations
 for LEP2 Physics}, CERN 2000-009 (2000), eds. S.~Jadach, G.~Passarino 
 and R.~Pittau, p. 1, hep-ph/0005309.
\bibitem{exca}
 F.~A.~Berends, R.~Pittau and R.~Kleiss,
 Comput.\ Phys.\ Commun.\  {\bf 85}, 437 (1995) and
 Nucl.\ Phys.\  {\bf B424}, 308 (1994).
\bibitem{hep} F.~A.~Berends, C.~G.~Papadopoulos and R.~Pittau, 
hep-ph/0002249 and ref.~\cite{wshop}, sections 5.1 and 6.2.
\bibitem{helac} A.~Kanaki and C~.G.~Papadopoulos, hep-ph/0002082. 
\bibitem{wopt}
R.~Kleiss and R.~Pittau,
Comput.\ Phys.\ Commun.\  {\bf 83} (1994) 141.
\bibitem{isr} 
E.~A.~Kuraev and V.~S.~Fadin, Yad.\ Fiz.\ {\bf 41}, 753 (1985)  
[Sov.\ J.\ Nucl.\ Phys.\ {\bf 41}, 466 (1985)];\\
G.~Altarelli and G.~Martinelli, in 
Physics al LEP, CERN-Yellow Report 86-06, eds. 
J. Ellis and R. Peccei (CERN, Geneva, February 1986);\\
O.~Nicrosini and L.~Trentadue, Phys. Lett. {\bf B196}, 551 (1987);\\
F.~A.~Berends, G.~Burgers and W.~L.~van Neerven,
Nucl.\ Phys.\ {\bf B297}, 429 (1988) and {\bf B304}, 921E (1988).
\bibitem{ptsf} 
G.~Montagna, M.~Moretti, O.~Nicrosini and F.~Piccinini,
Nucl.\ Phys.\  {\bf B541}, 31 (1999);\\
O.~Nicrosini and L.~Trentadue, Nucl.\ Phys.\ {\bf B318}, 1 (1989) and
Phys.\  Lett.\  {\bf B231}, 487 (1989).
\bibitem{ptsf1} 
F.~A.~Berends and R.~Kleiss, Nucl.\ Phys.\  
{\bf B260}, 32 (1985) and  {\bf B178}, 141 (1981).
\bibitem{sbhab}
W.~Beenakker, F.~A.~Berends and S.~C.~van der Marck, 
Nucl.\ Phys.\  {\bf B349}, 323 (1991).
\bibitem{isr2}
Y.~Kurihara, J.~Fujimoto, Y.~Shimizu, K.~Kato, K.~Tobimatsu and T.~Munehisa,
Prog.\ Theor.\ Phys.\  {\bf 103}, 1199 (2000).
\bibitem{gge}
F.~A.~Berends, P.~H.~Daverveldt and R.~Kleiss,
Nucl.\ Phys.\ {\bf B253}, 421 (1985).
\bibitem{isr3}
G.~Montagna, M.~Moretti, O.~Nicrosini,~A.~Pallavicini and F.~Piccinini,
hep-ph/0005121 
\bibitem{bfact} S.~Jadach, W.~Placzek, M.~Skrzypek and B.~F.~L.~Ward,
Phys.\ Rev.\  {\bf D54}, 5434 (1996).
\bibitem{passa1} G.~Passarino, Nucl.\ Phys.\  {\bf B574}, 451 (2000) and
{\bf B578}, 3 (2000).
\bibitem{passa2} W.~Beenakker et al., Nucl.\ Phys.\ {\bf B500}, 255 (1997).
\bibitem{passa3} E.~Argyres et al., Phys.\  Lett.\ {\bf B358},  339 (1995).
\bibitem{ballestrero} 
See the contribution by {\tt WPHACT} to ref.~\cite{wshop},
and E.~Accomando, A.~Ballestrero and E.~Maina, 
Phys.\ Lett.\  {\bf B479}, 209 (2000)
\bibitem{bbc} W.~Beenakker, F.~A.~Berends and A.~P.~Chapovsky,
Nucl.\ Phys.\  {\bf B573}, 503 (2000).
\bibitem{jeger}
S.~Eidelman and F.~Jegerlehner, Z.\ Phys.\  {\bf C67} (1995) 585.
\end{thebibliography}
\end{document}